\documentclass[10pt,aps,prb,twocolumn,floatfix,floats,showpacs,groupedaddress,raggedbottom]{revtex4-1}
\usepackage{graphicx,latexsym}
\usepackage{dcolumn}
\usepackage{amssymb,amsmath,bm}
\pdfoutput=1

\begin{document}

\title{Excitation of radial collective modes
       in a quantum dot: Beyond linear response}

\author{Vidar Gudmundsson}
\email{vidar@hi.is}
\author{Sigtryggur Hauksson}
\author{Arni Johnsen}
\author{Gilbert Reinisch}
\affiliation{Science Institute, University of Iceland, Dunhaga 3,
        IS-107 Reykjavik, Iceland}
\author{Andrei Manolescu}
\affiliation{School of Science and Engineering, Reykjavik University, Menntavegur 1, IS-101 Reykjavik, Iceland}
\author{Christophe Besse}
\affiliation{Institut de Math{\'e}matiques de Toulouse UMR 5219, Universit{\'e} de Toulouse; CNRS
             UPS IMT, F-31062 Toulouse Cedex 9, France}
\author{Guillaume Dujardin}
\affiliation{Inria Lille Nord-Europe et Laboratoire Paul Painlev{\'e} UMR CNRS 8524, 40 Avenue Halley,
             59650 Villeneuve d'Ascq Cedex, France}
%

\begin{abstract}
We compare the response of five different models of two interacting electrons in a quantum dot
to an external short lived radial excitation that is strong enough to excite the system
well beyond the linear response regime. The models considered describe the Coulomb interaction
between the electrons in different ways ranging from mean-field approaches to configuration 
interaction (CI) models, where the two-electron Hamiltonian is diagonalized in a large 
truncated Fock space. The radially symmetric excitation is selected in order 
to severely put to test the different approaches to describe the interaction and
correlations of an electron system in a nonequilibrium state.  
As can be expected for the case of only two electrons none of the mean-field models can in 
full details reproduce the results obtained by the CI model. Nonetheless, some linear and
nonlinear characteristics are reproduced reasonably well. 
All the models show activation of an increasing number of collective modes as the strength
of the excitation is increased. By varying slightly the confinement potential of the dot
we observe how sensitive the properties of the excitation spectrum are to the Coulomb interaction
and its correlation effects. In order to approach closer the question of nonlinearity we solve
one of the mean-field models directly in a nonlinear fashion without resorting to iterations.      

\end{abstract}

\pacs{73.23.-b, 78.67.-n, 42.50.Pq, 73.21.Hb}

\maketitle
\section{introduction}
Far-infrared spectroscopy and transport measurements were from early on 
used to investigate the electronic structure\cite{Reimann02:1283} of quantum dots of various
types. Far-infrared spectroscopy of arrays of quantum 
dots\cite{Demel90:788a,Shikin91:11903a} turned out to be rather
insensitive to the exact form of the interaction between the electrons. 
The reason being that most arrays of quantum dots resulted in almost
parabolic confinement of electrons to individual dots in the low energy regime. 
Soon it was realized that an exact symmetry condition, known as the  
the extended Kohn's theorem\cite{Kohn61:1242}
is valid for such systems as long as each  
dot is much smaller than the wavelength of the dipole radiation, and results in a pure 
center-of-mass motion of the electrons in each dot, independent of the number of
electrons and the nature of the interaction between them. 
Signatures of deviations from the parabolic confinement where soon discovered in
experimental results and interpreted with model calculations based on various 
approaches to linear response.\cite{Pfannkuche91:13132,Gudmundsson91:12098,Pfannkuche93:6} 
The Coulomb blockade helped
guaranteeing a definite number of electrons in each quantum dot homogeneously in the large arrays 
that were necessary to allow measurement of the weak FIR absorption signal.
Deviations from the parabolic confinement of electrons in quantum dots 
lead to the excitation of internal collective modes that can cause splitting of the
upper plasmonic branch and make visible the classical Bernstein\cite{Bernstein58:10} 
modes.\cite{Gudmundsson95:17744,Krahne01:195303} 
In the lower plasmonic branch they lead to weak oscillations caused by filling factor dependent 
screening properties.\cite{Bollweg96:2774,Darnhofer96:591}   

Resonant Raman scattering has been applied to quantum dots to analyze ``single-electron''
excitations and collective modes with monopole, dipole, or quadrupole symmetry 
($\Delta M=0,\pm1,\pm2$).\cite{Steinebach99:10240,Steinebach00:15600} 
As the monopole collective oscillations are excitations that can be exclusively 
described by internal relative coordinates one would expect them to be more influenced by 
the Coulomb interaction between the electrons than the dipole excitations that have to be
described by relative and center-of-mass coordinates, or purely by the latter ones
when the Kohn theorem holds.\cite{Kohn61:1242} 
The $\Delta M=0$ collective mode among 
others was measured by a very different method and calculated for a confined two-dimensional 
electron system in the classical regime on the surface of liquid Helium.\cite{PhysRevLett.54.1710}

In the far-infrared and the Raman measurements of arrays of dots the excitation has
always been weak and some version of linear response has been an adequate approach to
interpret the experimental results. All the same, curiosity has driven theoretical groups 
into questioning how the electron system in a quantum dot would respond as the linear
regime is surpassed and a strong excitation would pump energy into the 
system.\cite{Puente01:235324,Gudmundsson03:161301,Gudmundsson03:0304571}   
These studies have been undertaken with some kind of a mean-field model to incorporate the 
Coulomb interaction between the electrons. 
Here, we will explore this nonlinear excitation regime with a model built on exact numerical
diagonalization or configuration interaction (CI)\cite{Pfannkuche93:6} and compare the results
with the predictions of three different mean-field approaches, and a time-dependent Hubbard model.
Besides the question of what happens in the nonlinear regime, we want to see how close to the 
exact results the mean-field models can come for only two electrons in the dot, a regime that 
is indeed challenging for mean-field approaches which in general are more appropriate 
for a higher number of electrons.
We will address issues of nonlinear behavior. What do we classify as nonlinear behavior?
Can we see it emerging in an exact model? 
How, and when is it inherent in a mean-field approach?

\section{Short excitation in the THz regime}
In order to describe the response to an excitation of arbitrary strength 
we will follow the time-evolution of the system by methods that are
appropriate to each model. 
At $t=0$ the quantum dot is radiated by a short THz pulse  
\begin{eqnarray} 
      W(t) &=& V_t r^{|N_p|}\cos{(N_p\phi)}\exp{(-sr^2-\Gamma t)}\nonumber\\ 
           &{\ }&\sin{(\omega_1t)}\sin{(\omega t)}\theta (\pi -\omega_1t),
\label{Wt}
\end{eqnarray}
where $\theta$ is the Heaviside step function. For the purpose of making the
response strongly dependent on the Coulomb interaction between the electrons
we select the monopole or the breathing mode with $N_p=0$. It should be kept in
mind that this short excitation pulse perturbs the system in a wide frequency range.

The quantum dot will have a parabolic confinement potential 
\begin{equation}
      V_{\mathrm{par}}(r)=\frac{1}{2}m^*\omega_0^2r^2,
\label{Vpar}
\end{equation}
with $\hbar\omega_0 = 3.37$ meV. In addition, we will sometimes add a small
potential hill in the center of the dot
\begin{equation}
      V_{\mathrm{c}}(r) = V_0\exp{(-\gamma r^2)},
\label{Vc}
\end{equation}
with $V_0 = 3.0$ meV, and $a^2\gamma = 1.0$, where $a = \sqrt{\hbar/(m^*\omega_0)}$ 
is the characteristic length scale for the parabolic confinement. We will be assuming
GaAs parameters here with $m^*=0.067m_e$ and a dielectric constant $\kappa = 12.4$.
If we select $sa^2 = 0.8$, $\hbar\omega_1 = 0.658$ meV,  $\hbar\omega = 2.63$ meV,
and $\Gamma =2.0$ ps$^{-1}$,
then the initial pulse of duration approximately $3$ ps represents a spatial circular Gaussian
pulse rising from zero and vanishing after its amplitude gets negative. 
The system is perturbed by a radial compression followed by a slight radial expansion
and then left to oscillate freely about the equilibrium point. The system will be kicked
out of equilibrium and the time-evolution has to be described accordingly for each model.

The reason for adding the central hill (\ref{Vc}) to the quantum dot is to avoid
any special symmetry that could result from the parabolic confinement (\ref{Vpar}).

\section{Time-evolution of quantum dot Helium with a DFT interaction}
The details of a density functional theoretical (DFT) approach to the model
used to describe nonadiabatic excitation of electrons in a quantum
ring or dots in an external magnetic field has been published 
earlier.\cite{Gudmundsson03:161301,Gudmundsson03:0304571}
Here, we will use the model for a vanishing external magnetic field and 
properly make clear the difference in the calculation of the time-evolution of this 
mean-field model to the CI model. To accomplish this we need to list few steps.  

\begin{figure}[htbq]
      \includegraphics[width=0.42\textwidth,angle=0]{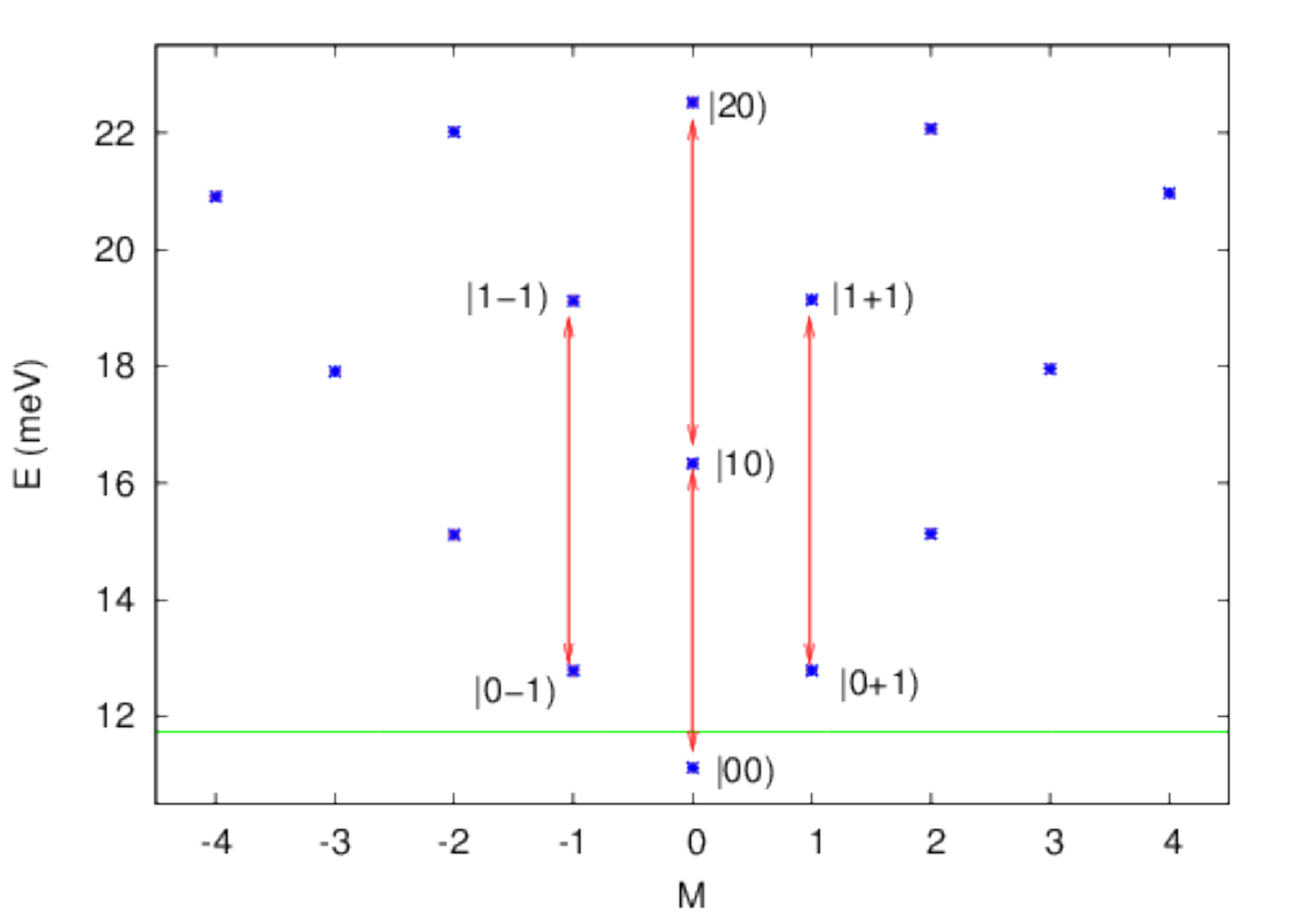}
      \caption{(Color online) The effective single-electron energy spectrum
               for the DFT-version of the model of two electrons in a
               parabolic quantum dot with a small central hill (\ref{Vc})
               as a function of the quantum number of angular momentum $M$.
               The chemical potential, $\mu$, needed to have two electrons
               in the dot is indicated by a solid green horizontal line.  
               $V_0=3$ meV, $T=0.1$ K.}
      \label{E-rof-dft}
\end{figure}

The ``single-electron'' energy spectrum of the model is presented in 
Fig.\ \ref{E-rof-dft} at temperature $T=0.1$ K and for a small hill (\ref{Vc}) placed
in the center of the system.
The finite, but small temperature is used to stabilize the iteration process
used to solve the DFT model. The chemical potential $\mu$ needed to have
two electrons in the ground state of the system is indicated in the figure
by a horizontal green line.
The calculation is a ``grid-free'' approach utilizing the eigenstates of the
noninteracting system as a functional basis $\{|nM\rangle\}$. The interacting
states $|\alpha )$ can not be assigned a definite quantum number $n$ and
$M$, but as the system is circularly symmetric here, by comparing the location 
in the energy spectrum and by checking the leading
contribution to the interacting states we allow ourselves to assign, for educational
purposes, the quantum numbers shown in Fig.\ \ref{E-rof-dft}. The central hill (\ref{Vc})
and the Coulomb interaction raise the energy of the states with high $M=0$ contribution.

To calculate the time-evolution of the system kicked out off equilibrium
by the perturbing pulse (\ref{Wt}) we use the Liouville-von Neumann equation for the density operator 
\begin{equation}
      i\hbar \frac{d}{dt}{\rho}(t) = [H + W(t),\rho (t)],
\label{L-vN-dft}
\end{equation}
represented in the noninteracting basis $\{|n,M\rangle\}$.  The structure of this equation is 
inconvenient for numerical evaluation so we resort instead to the time-evolution operator $T$, 
defined by $\rho (t) = T(t)\rho_0T^+(t)$, which has the simpler equation of motion
\begin{eqnarray}
      i\hbar\dot T(t)   &=& H(t)T(t)\nonumber\\    
     -i\hbar\dot T^+(t) &=& T^+(t)H(t).
\label{Teq}     
\end{eqnarray}
The single-electron basis is truncated after tests for convergence of the time-evolution 
with the parameters used here.  We discretize time and use the Crank-Nicholson algorithm 
for the time-integration with the initial condition, $T(0)=1$.  

\begin{figure}[htbq]
      \includegraphics[width=0.42\textwidth,angle=0]{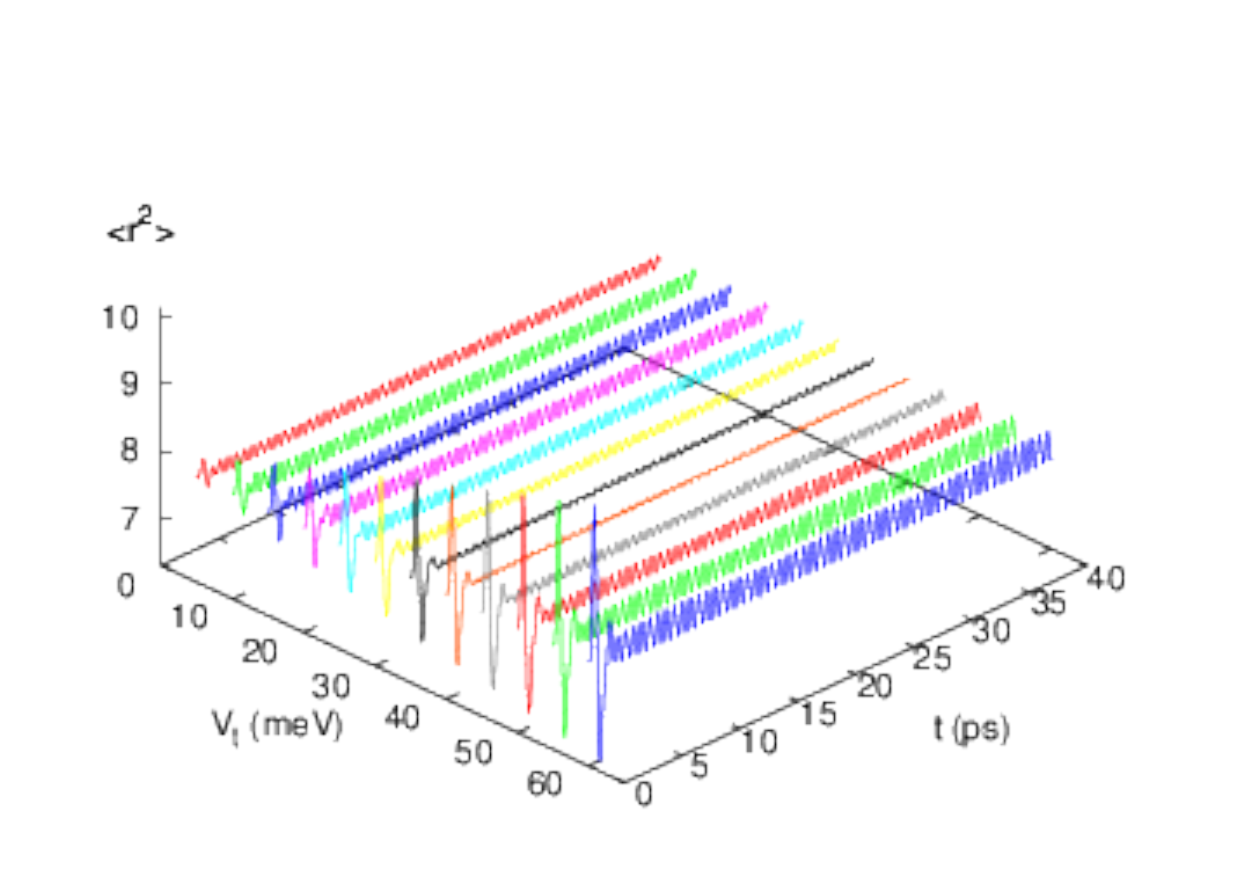}
      \caption{(Color online) The time-evolution of the expectation value
              $\langle r^2\rangle$ as function of the strength of the initial
              perturbation pulse, $V_t$, for the DFT-version of the model of
              two electrons in a quantum dot. $V_0=3$ meV, $T=0.1$ K.}
      \label{r2-dft-T01-hill}
\end{figure}

\begin{figure}[htbq]
      \includegraphics[width=0.46\textwidth,angle=0]{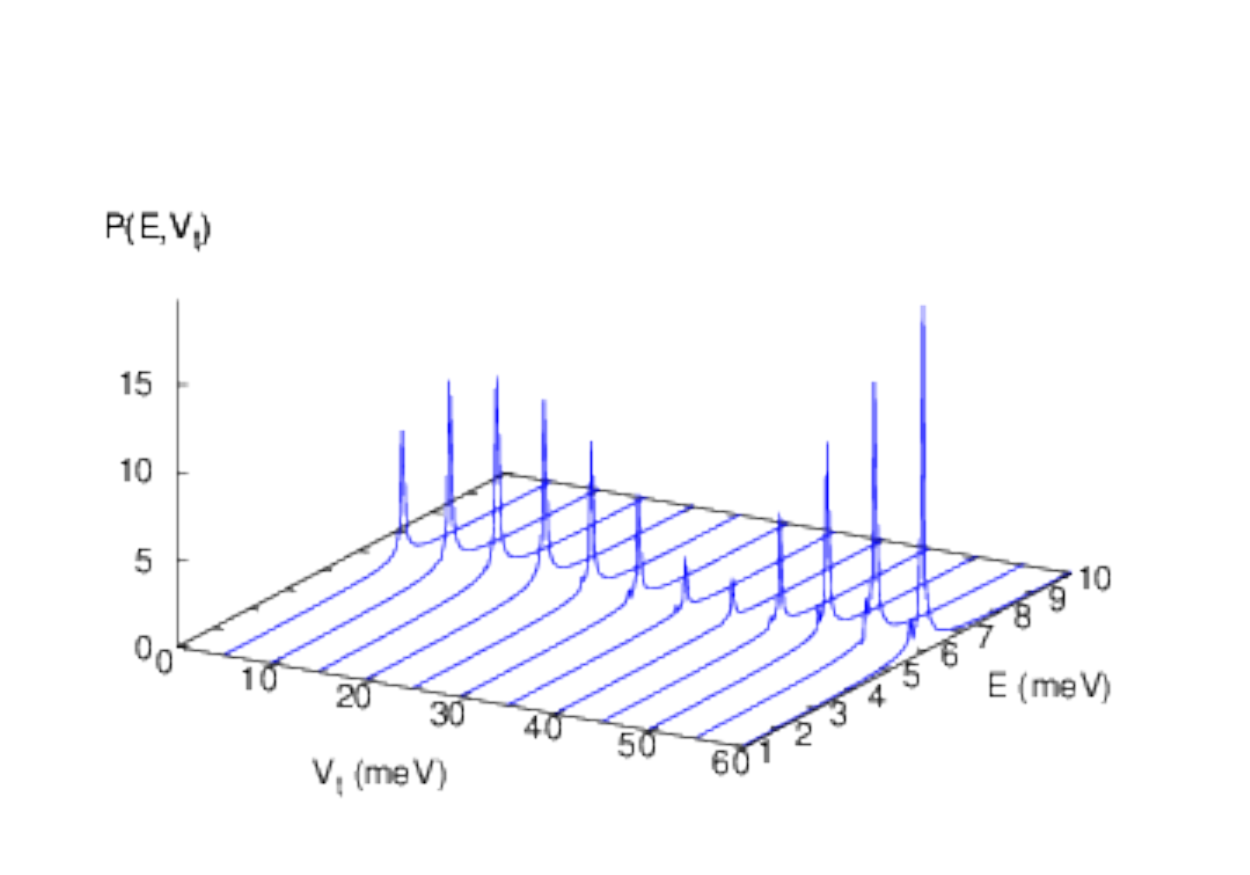}
      \includegraphics[width=0.46\textwidth,angle=0]{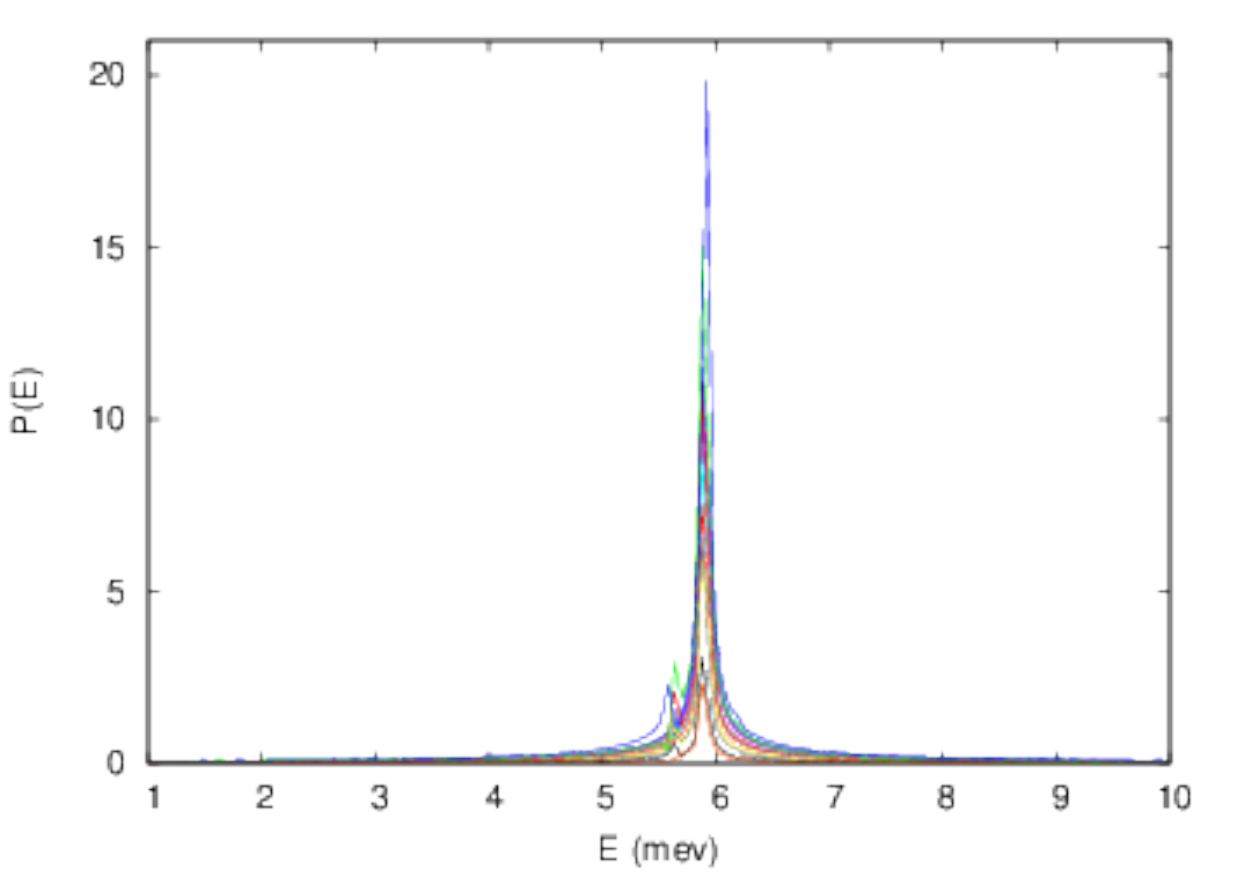}
      \caption{(Color online) The Fourier power spectrum for the time-evolution of
              $\langle r^2\rangle$ for the DFT-version of the model of two-electrons
              in a quantum dot. The lower panel is a side view to demonstrate the 
              stability in frequency for different values of excitation $V_t$. 
              $V_0=3$ meV, $T=0.1$ K.}
      \label{FFT-r2-dft-T01-hill}
\end{figure}

The circular symmetry of the confinement potential (Eq.'s (\ref{Vpar}) and (\ref{Vc})) and the
excitation pulse (\ref{Wt}) suggest the mean value of the radius squared to be an
ideal observable to be analyzed. In Fig.\ \ref{r2-dft-T01-hill} we show $\langle r^2\rangle$
as function of time $t$ and the strength of the perturbing pulse $V_t$.
We see already in Fig.\ \ref{r2-dft-T01-hill} that the amplitude of the response to the
initial perturbation (\ref{Wt}) is nonlinear. To analyze this better we show the Fourier
transform in Fig.\ \ref{FFT-r2-dft-T01-hill}(a), where we indeed see a local minimum around
$V_t\approx 35-40$ meV.   

\begin{figure}[htbq]
      \includegraphics[width=0.42\textwidth,angle=0]{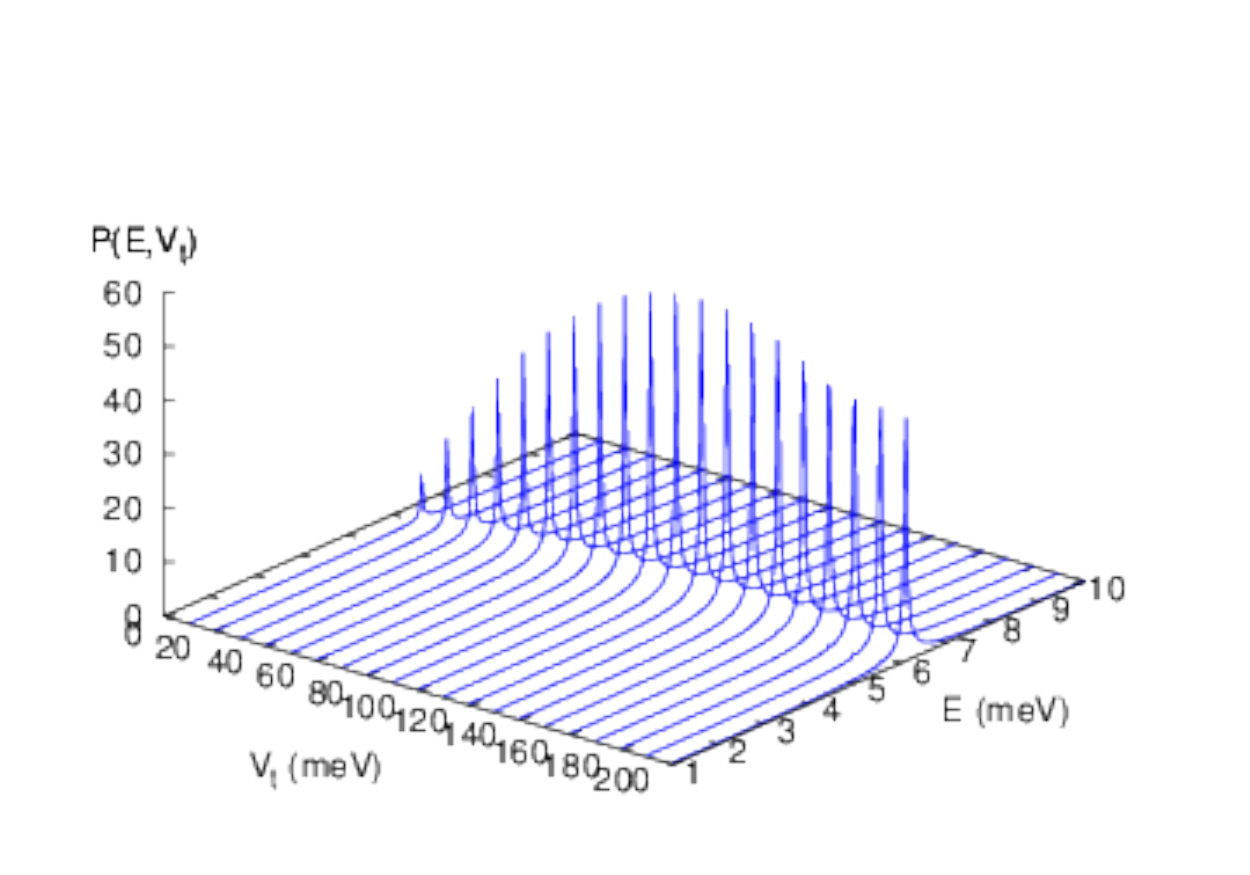}
      \includegraphics[width=0.42\textwidth,angle=0]{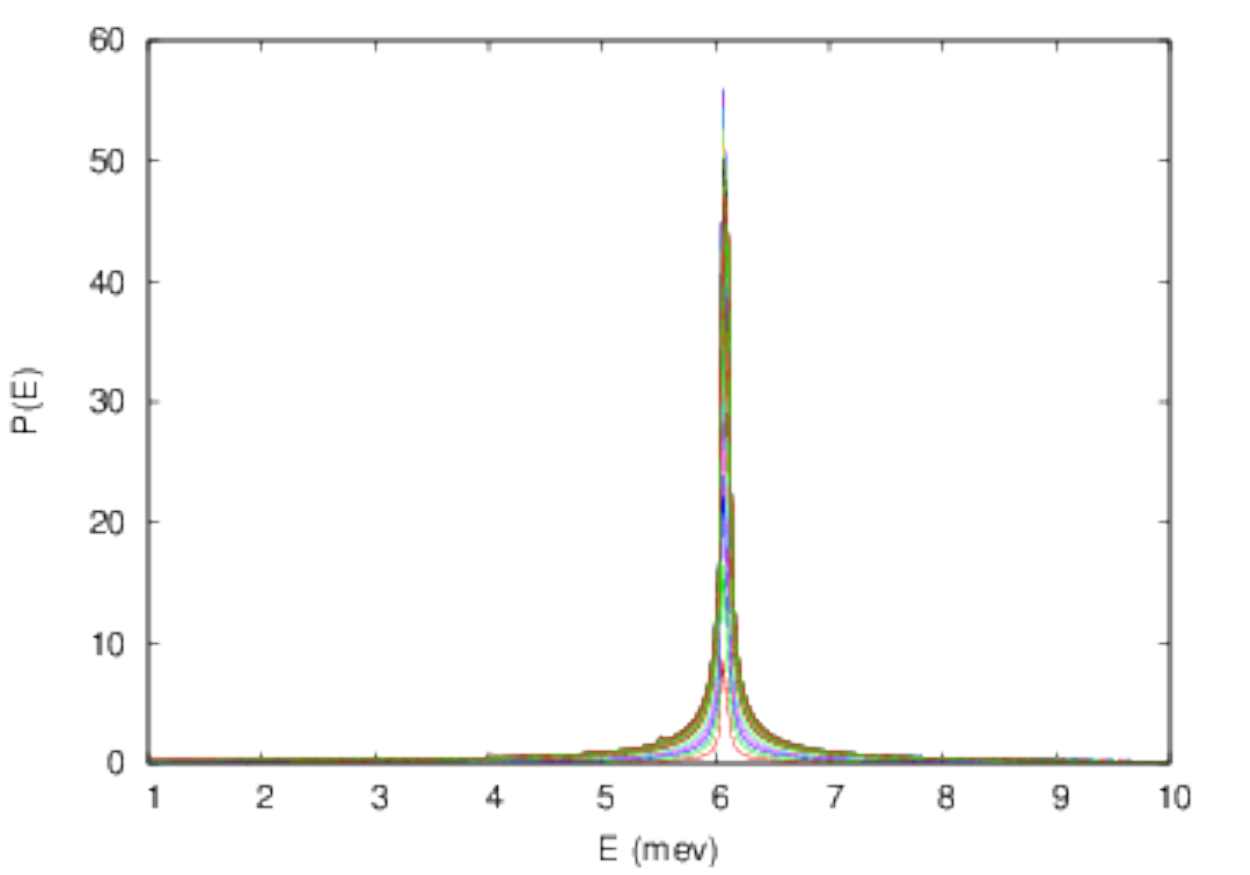}
      \caption{(Color online) The Fourier power spectrum as a function of energy and
              perturbation strength $V_t$ for the Hartree Approximation.
              $V_0=3.0$ meV, $T=0.1$ K.}
      \label{FFT-r2-hartree-T01} 
\end{figure}

\begin{figure}[htbq]
      \includegraphics[width=0.23\textwidth,angle=0,bb=0 0 160 245,clip]{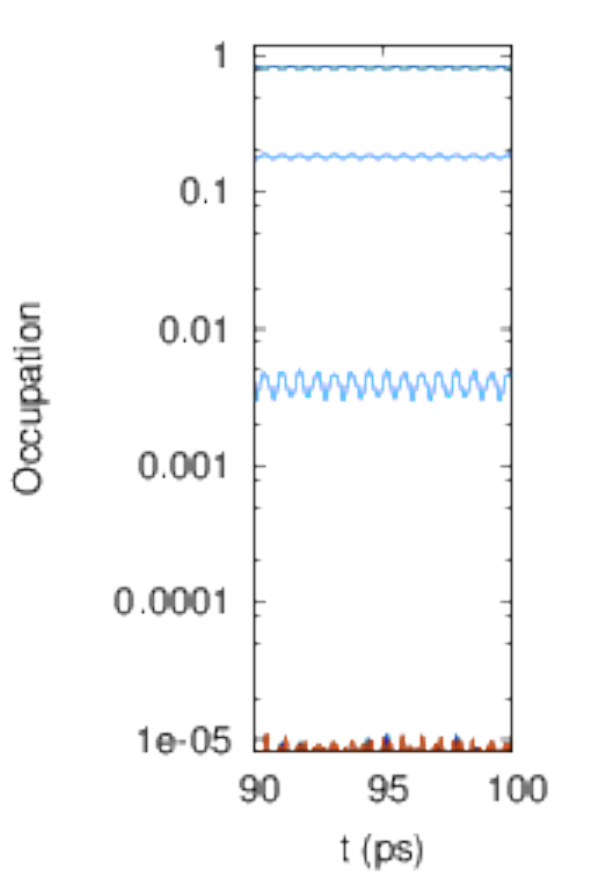}
      \includegraphics[width=0.23\textwidth,angle=0,bb=0 0 160 245,clip]{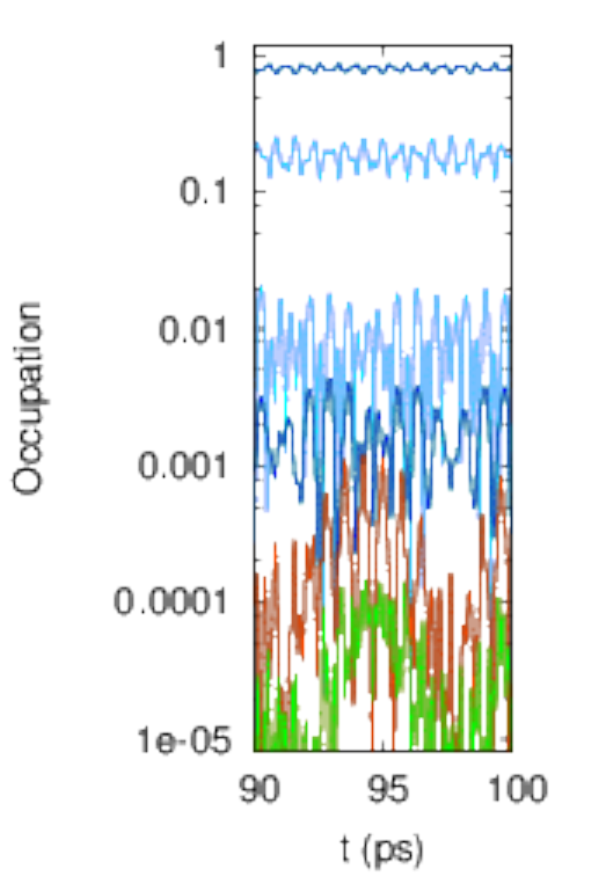}
      \caption{(Color online) Time-dependent occupation of effective single-electron states (of the noninteracting
              basis ${|nM\rangle}$) for the HA model with a central hill (\ref{Vc}) for
              $V_t=10.0$ meV (left panel), and $V_t=200.0$ meV (right panel). $V_0=3.0$ meV, $T=0.1$ K.}
      \label{Occupation-T}
\end{figure}

Curiously enough, this local minimum can not be seen in the results if we turn off
the exchange and the correlation functionals in the DFT model, i.e.\ if we use a
Hartree approximation (HA) for the Coulomb interaction, see Fig.\ \ref{FFT-r2-hartree-T01}.
For a later discussion we note here that the time-dependent HA calculations for the 
present parameters are much more stable then the DFT version. We are thus able to go to
higher values of $V_t$ and observe the time-evolution for longer time resulting in more
accurate Fourier transforms. In the DFT or the HA model the part of the Hamiltonian describing
the effective Coulomb interaction remains time-dependent at all times, even after the 
initial perturbing pulse has vanished, since the local effective potential depends on the 
electron density which is oscillating in time. It is thus of no surprise that in these
mean-field models the occupation, the diagonal elements of the density matrix (\ref{L-vN-dft}),
remain time-dependent as can be seen in Fig.\ \ref{Occupation-T}. 
This time-dependence of the occupation and the effective interaction will be in
contrast to what happens in the CI calculation described below. 

In a real system, an open system, the oscillations will be damped by phonon 
interactions\cite{PhysRevB.75.125324} or photons.\cite{PhysRevB.87.035314}
In the far-infrared regime the radiation time scale is much longer than the 
100 ps during which we follow the evolution of the system here.  

\begin{figure}[htbq]
      \includegraphics[width=0.21\textwidth,angle=0,bb=42 36 313 207,clip]{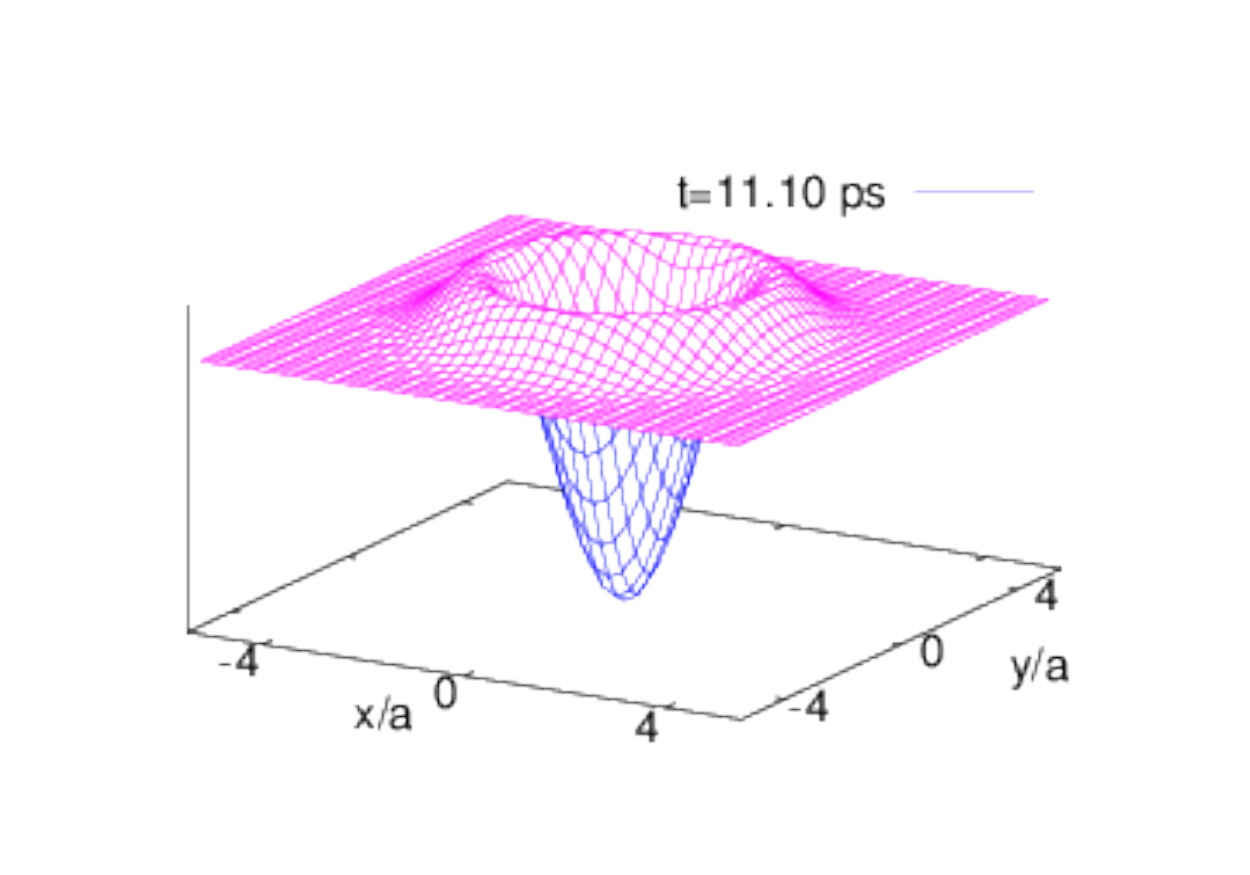}
      \includegraphics[width=0.21\textwidth,angle=0,bb=42 36 313 207,clip]{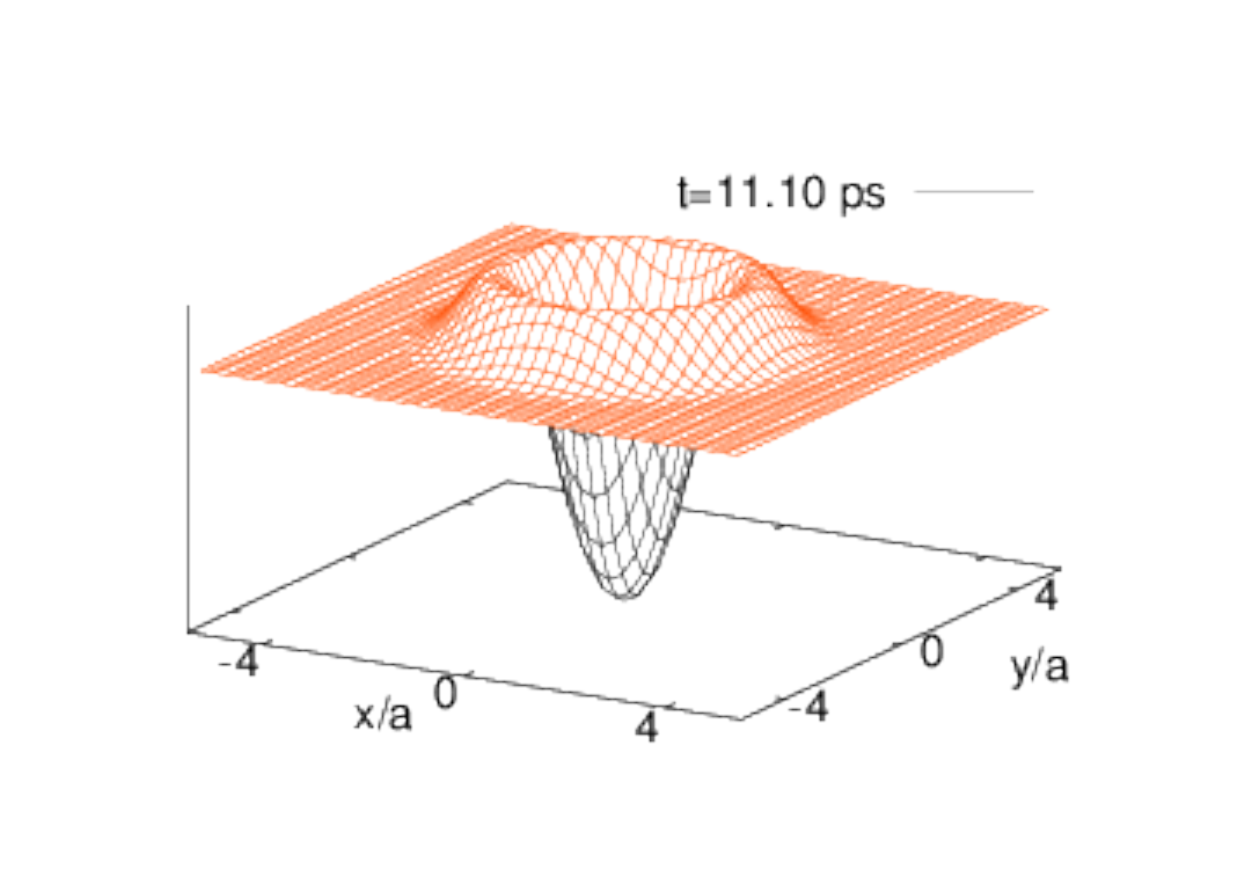}\\
      \includegraphics[width=0.21\textwidth,angle=0,bb=42 36 313 207,clip]{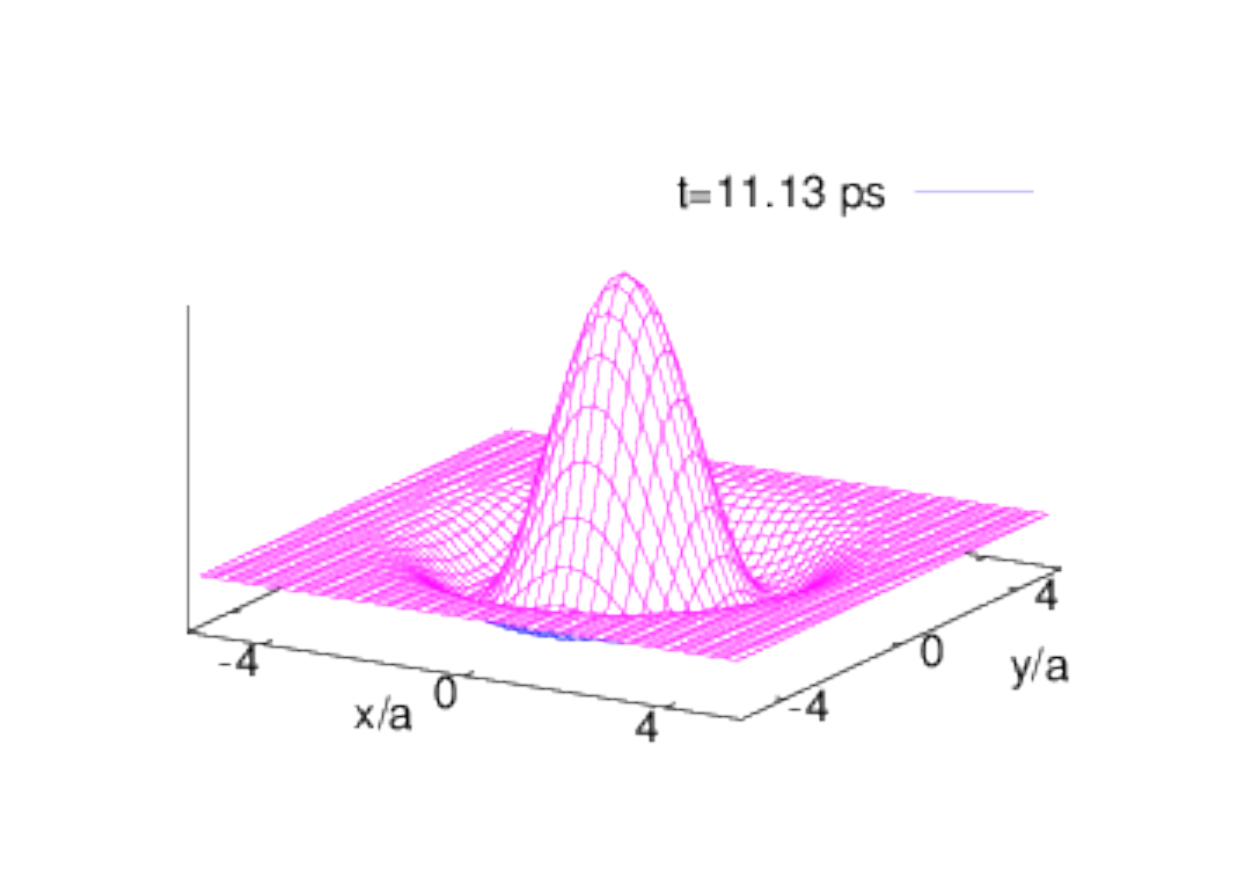}
      \includegraphics[width=0.21\textwidth,angle=0,bb=42 36 313 207,clip]{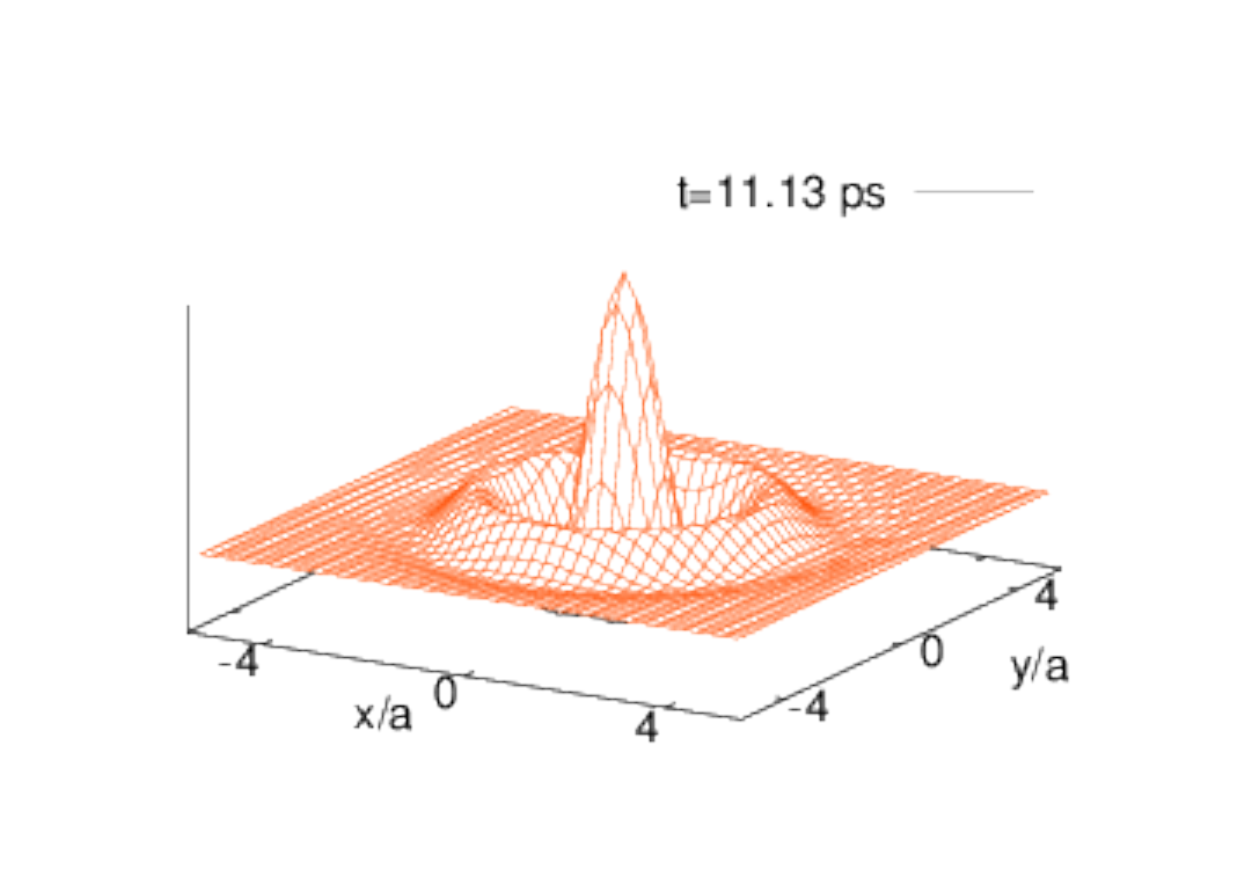}\\
      \includegraphics[width=0.21\textwidth,angle=0,bb=42 36 313 207,clip]{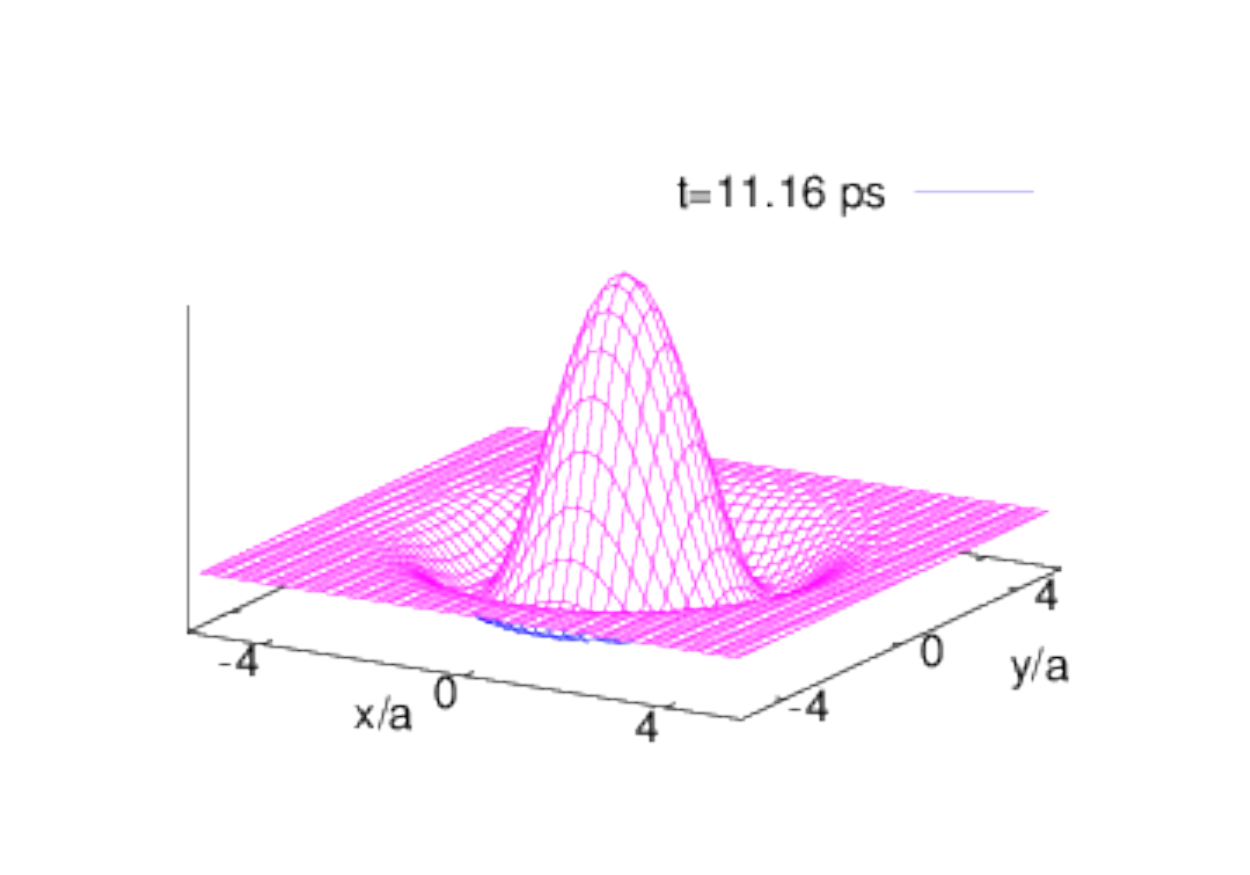}
      \includegraphics[width=0.21\textwidth,angle=0,bb=42 36 313 207,clip]{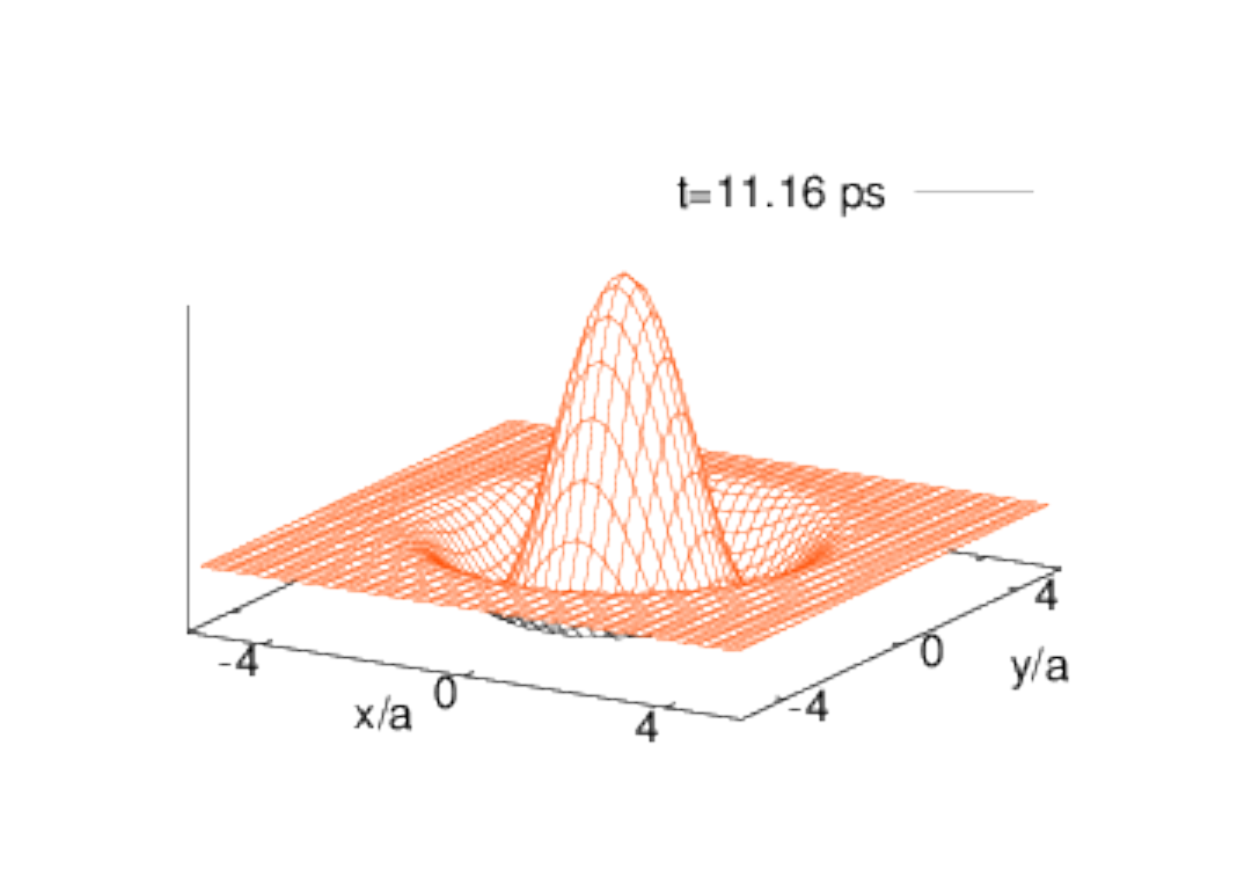}\\
      \includegraphics[width=0.21\textwidth,angle=0,bb=42 36 313 207,clip]{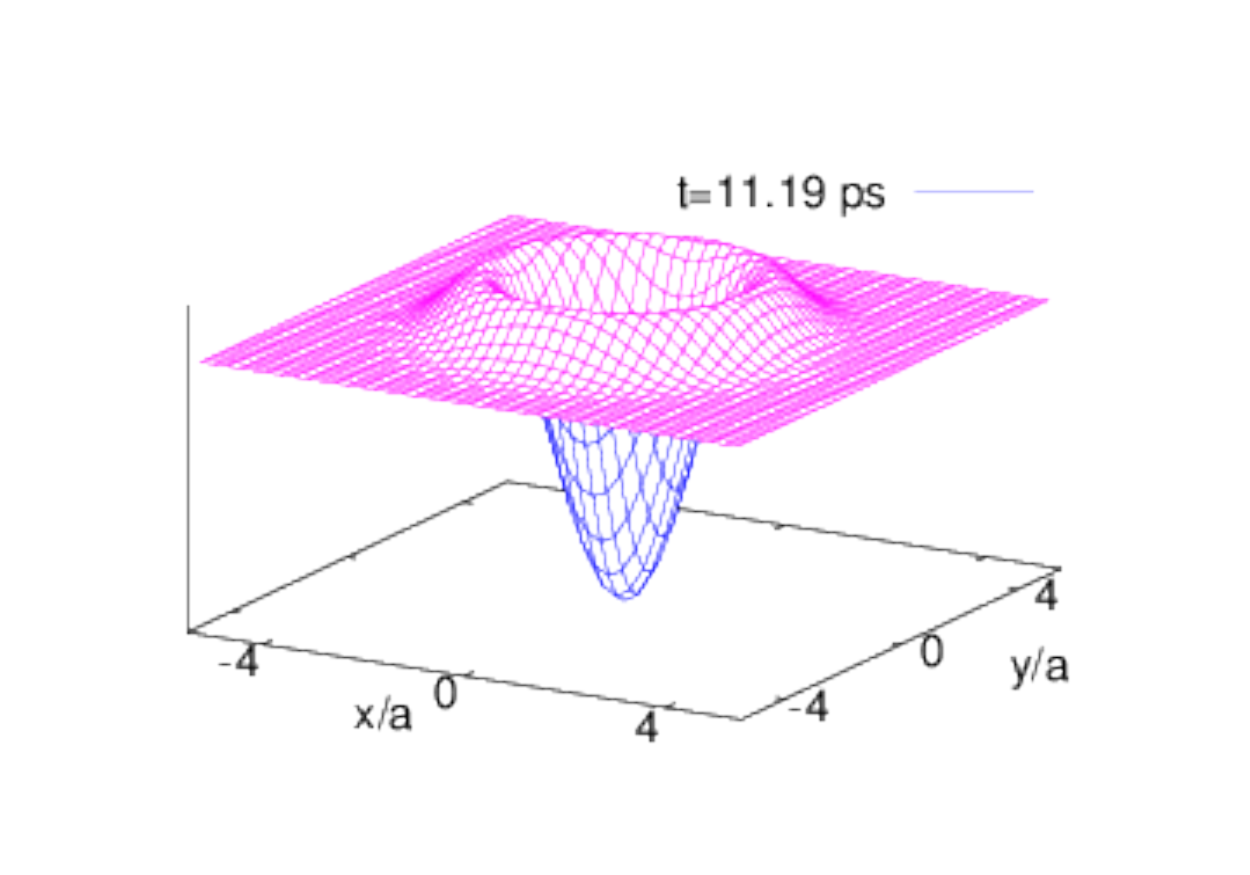}
      \includegraphics[width=0.21\textwidth,angle=0,bb=42 36 313 207,clip]{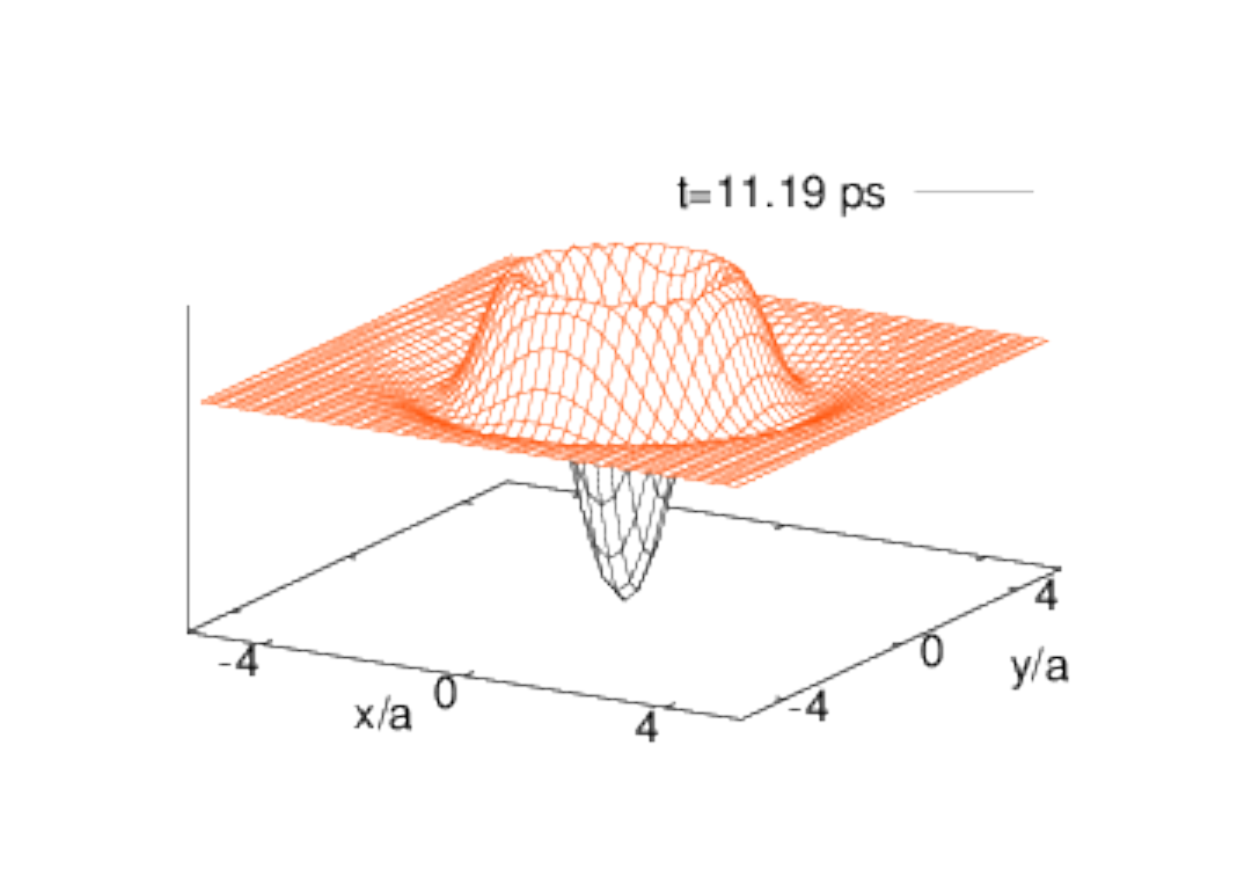}
      \caption{(Color online) The induced density $\delta n(r,t)=n(r,t)-n(r,0)$ within one period
              for the DFT model for $V_t=5$ meV (left), and $V_t=35$ meV (right). 
              $V_0=3.0$ meV, $T=0.1$ K.}
      \label{dn-dft-050-350}
\end{figure}

It is possible to construct the time-dependent induced density, $\delta n(r,t)=n(r,t)-n(r,0)$, 
for the oscillations in the system in the hope to monitor the modes being occupied for different values of
$V_t$. In Fig.\ \ref{dn-dft-050-350} we see the induced density for the DFT model over approximately one
oscillation for $V_t=5$ and $35$ meV. It is clear that for the higher value of excitation
a second oscillation mode is superimposed on the fundamental mode visible for $V_t=5$ meV.  
For still higher excitation this becomes even more apparent. In Fig.\ \ref{E-rof-dft} the main
``single-electron'' contribution to this collective oscillation is indicated by an arrow between
$|00)$ and $|10)$. Higher excitation brings in a mixing from the $|10)$ to $|20)$ transition, and
higher temperature would activate transitions from $|0-1)$ to $|1-1)$, and from $|01)$ to $|11)$. 

\section{Time-evolution of a quantum dot Helium described by a nonlinear
         Schr{\"o}dinger-Poisson equation}
We will consider one more variant of a mean-field model for the two
Coulomb interacting electrons in the quantum dot. This model could be considered a 
version of the HA for a special case, but we investigate it here for a different
reason. It allows for the application of a nonlinear solution method to be described 
at the end of this section.\cite{Reinisch11:699,Reinisch12:902}

Consider a $S=0$  electron pair located at $z_{1,2}=x_{1,2}+iy_{1,2}$  in the $x-y$
plane and confined by the 2D parabolic potential (\ref{Vpar}). 
Since their spins are opposite, both electrons can stay, as fermions, in the same orbital state $\psi$. 
Moreover, they obey a pair orbital symmetry. Therefore the simplest 
two-electron wavefunction $\Psi_{\mathrm{pair}}(z_1,z_2,t)$ is
\begin{equation}
\label{eq-orbital2el}
      \Psi_{\mathrm{pair}}(z_1,z_2,t)=\psi(z_1,t)  \psi(z_2,t), 
\end{equation}
where $|\Psi_{\mathrm{pair}}(z_1,z_2,t)|^2$ is the probability density to find at time $t$ either electron at
$z_i$ while the other is at
$z_j$ ($i\neq j = 1,\,2$).  Therefore, the normalization condition reads
\begin{equation}
\label{eq-norm}
      \int d^2z_1 d^2z_2 |\Psi_{\mathrm{pair}}(z_1,z_2,t)|^2 =
      \left[\int d^2z |\psi(z,t)|^2\right]^2=1.
\end{equation}
We assume that  $\psi(z,t)\equiv \psi(x,y,t)$  is a  
time-dependent nonlinear state defined by the following   
Schr{\"o}dinger-Poisson (SP) differential system
\begin{align}
\label{eq-Schroe_vraie}
      i\hbar \frac{\partial}{\partial t} \psi &= H \psi,\\
\label{eq-Poisson_vraie}
      \nabla^2 \Phi &= -2\pi{\cal N}\hbar\omega |\psi|^2,
\end{align}
where ${\cal N}$ is a  dimensionless
order parameter of the SP system
that defines the strength of the Coulomb repulsive interaction potential $\Phi$
between the particles in units of $\hbar\omega_0$ (in a loose sense, we call it the ``norm'': see 
below Eq.\ (\ref{eq-normu})). The 2D nonlinear Hamiltonian is defined by
\begin{equation}
\label{eq-H_vrai}
      H =-\frac{\hbar^2}{2m^*} \nabla^2 +\Phi(x,y,t) 
      + \frac{1}{2} m^* \omega_0^2 (x^2+y^2).
\end{equation}
Using the characteristic length $a$ of the parabolic confinement 
and its frequency $\omega_0$ we perform the following 
change of variables
\begin{equation}
\label{eq-change_var}
      X=\frac{x}{a};\enspace Y=\frac{y}{a};\enspace \tau=\omega t
      ;\enspace \psi =\sqrt{\frac{2m^*\omega_0}{\hbar{\cal N}}}u(X,Y,\tau).
\end{equation}
Accordingly, Eq.\ (\ref{eq-norm}) becomes
\begin{equation}
\label{eq-normu}
      \int |u(X,Y,\tau)|^2 dX dY ={\cal N},
\end{equation}
while the SP time-space differential system (\ref{eq-Schroe_vraie}-\ref{eq-H_vrai}) yields
\begin{equation}
\label{eq-Sdimless}
      i\frac{\partial}{\partial \tau}u +\nabla_{X,Y}^2 u - V u =0,
\end{equation}
\begin{equation}
\label{eq-Pdimless}
      \nabla_{X,Y}^2 V +|u|^2-1 =0,
\end{equation}
where $\nabla_{X,Y}$ operates on the new variables $X$ and $Y$.
The (time-dependent) effective mean-field dimensionless
potential experienced by the particles is
\begin{equation}
\label{eq-POTdimless}
      V= \frac{\frac{1}{2} m^* \omega^2 (x^2+y^2)+\Phi}{\hbar\omega_0}
      =\frac{1}{4}(X^2+Y^2)+\frac{\Phi}{\hbar\omega_0}.
\end{equation}
We wish to define the observable which allows comparison with
the previous sections. Labelling $\bar z=\frac{1}{2}(z_1+z_2)$, 
$\bar x=\frac{1}{2}(x_1+x_2)$, and $\bar y=\frac{1}{2}(y_1+y_2)$,
we have $\bar z \bar z^*={\bar x}^2+{\bar y}^2$ and therefore
\begin{equation}
\label{eq-MeanVal}
      \langle\langle  \bar z \bar z^* \rangle\rangle=
      \frac{1}{2}\Bigl[  \langle x^2  \rangle  + \langle y^2  \rangle +  
      \langle x  \rangle^2 + \langle y \rangle^2\Bigr],
\end{equation}
where for any observable $A$
\begin{equation}
\label{eq-MeanPair}
      \langle\langle  A\rangle\rangle=\int d^2z_1 d^2z_2 A |\Psi_{\mathrm{pair}}|^2 ,
\end{equation}
and
\begin{equation}
      \label{eq-MeanSP}
      \langle  A\rangle=\int dx dy A |\psi|^2,
\end{equation}
(cf.\ Eq.\ (\ref{eq-norm})). Obviously, $\sqrt{\langle\langle  \bar z \bar z^* \rangle\rangle}$
is a sound measure of the time-dependent extension of the system. In the dimensionless
variables (\ref{eq-change_var}), it reads
\begin{equation}
\label{eq-MeanSPreduced}
      R(\tau)= \frac{1}{\sqrt{2}}\Bigl[  \langle X^2  \rangle_u  + \langle Y^2  \rangle_u  +  
      \langle X  \rangle_u ^2 + \langle Y \rangle_u ^2\Bigr]^{\frac{1}{2}},
\end{equation}
where
\begin{equation}
\label{eq-MeanSPred}
      \langle  A\rangle_u=\frac{1}{{\cal N}} \int dX dY A |u|^2 ,
\end{equation}
(cf.\ Eq.\ (\ref{eq-normu})).

The solution of system (\ref{eq-Schroe_vraie}-\ref{eq-Poisson_vraie}) demands the
initial profile $\psi(x,y,0)$. For these means we use the radial symmetric ground
state of the time-independent system. The Poisson equation (\ref{eq-Poisson_vraie})
is two-dimensional here and would thus produce a logarithmic Green function for
homogeneous space instead of the $1/r$ three-dimensional that we should be using
since the electric field can not be confined to 2D even though the electrons can be.
But, we do accept this discrepancy for three reasons. First, the asymptotic behavior at
$r\sim 0$ is not so dissimilar though the logarithm represents a bit softer repulsion, 
and second, the long range behavior will not carry
much weight due to the parabolic confinement potential (\ref{Vpar}). Third, and most
important, the SP system (\ref{eq-Schroe_vraie}-\ref{eq-Poisson_vraie}) can be solved
directly to obtain a nonlinear solution.\cite{Reinisch11:699,Reinisch12:902}
Generally, for physical mean-field models, which are of course nonlinear, the traditional
method is to seek a solution by iteration. In case of the HA or the DFT model here, the 
effective interaction potential is calculated after an initial guess has been made for the 
wavefunctions. Then the new wavefunctions are sought by methods from linear algebra, and
the iterations are continued until convergence is reached. The wavefunctions will be
orthonormal. When the SP system is solved directly the wavefunctions are not in general
orthonormal. Besides convenience, the reason for the iteration method is the connection
of the Hartree and Hartree-approximations to higher order methods in many-body theory,
that can only be established in case of orthonormal solutions. 
The hope is that the iteration method supplies the nonlinear solution in this 
sense or a solution very close to it. In fact, the nonlinear solutions are almost 
orthonormal with some small discrepancy of the order of $1-5\%$.        

\begin{figure}[htbq]
      \includegraphics[width=0.46\textwidth,angle=0,bb=123 45 1411 722,clip]{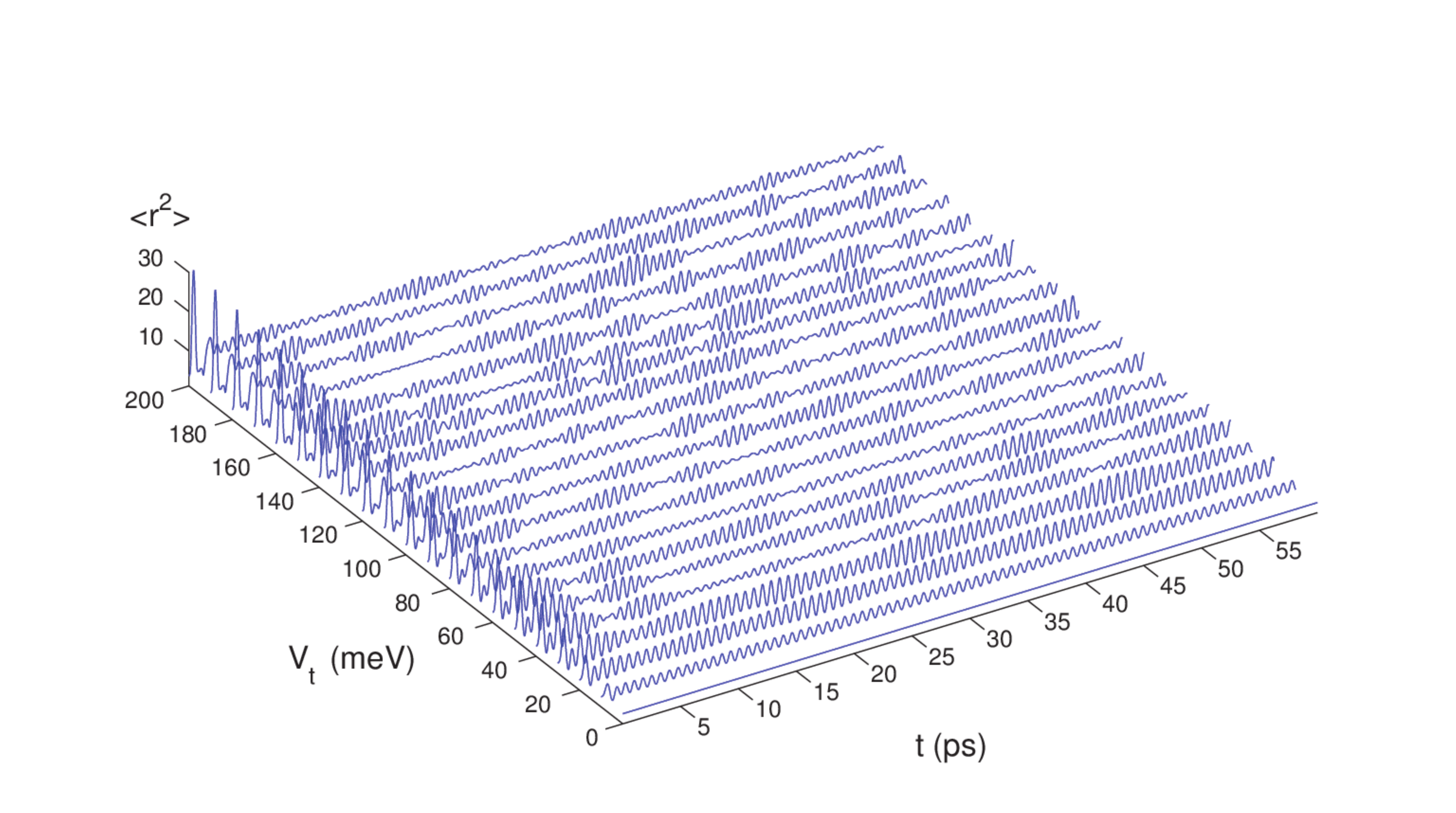}
      \includegraphics[width=0.46\textwidth,angle=0,bb=123 45 1411 722,clip]{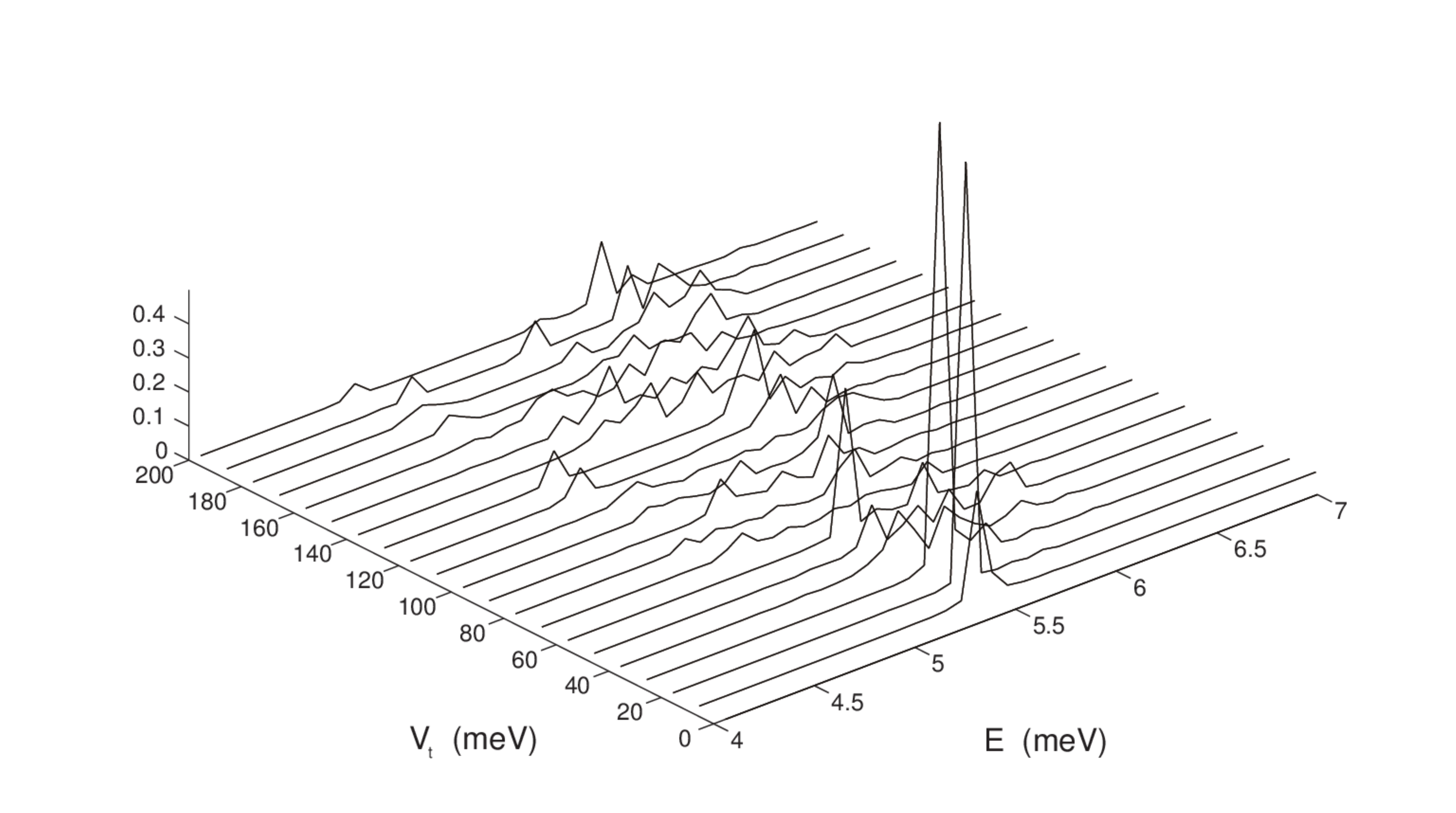}
      \caption{(Color online) The time-dependent expectation value of 
              $\langle r^2 \rangle$ and the corresponding Fourier power spectrum for the 
              Schr{\"o}dinger-Poisson model of the quantum dot without a central hill. $V_0=0$, $T=0$ K.}
      \label{FFT-Schr-Poisson-nohill}
\end{figure}

In Fig.\ \ref{FFT-Schr-Poisson-nohill} we show the results for the time-evolution of the expectation value 
$\langle\langle  \bar z \bar z^* \rangle\rangle$ and the corresponding Fourier
transform for the SP model without a central hill in the quantum dot.   
\begin{figure}[htbq]
      \includegraphics[width=0.46\textwidth,angle=0,bb=123 45 1411 722,clip]{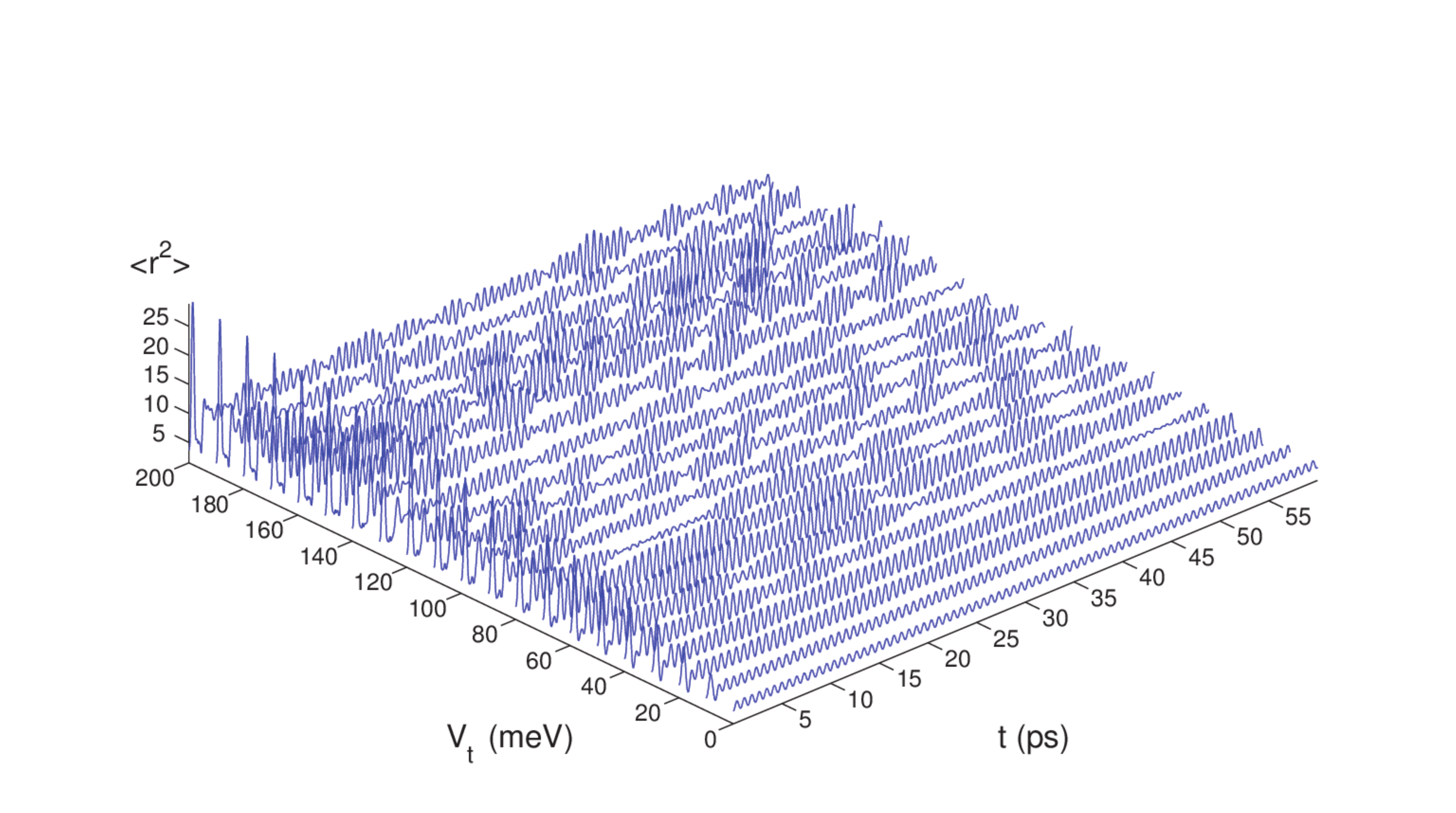}
      \includegraphics[width=0.46\textwidth,angle=0,bb=123 45 1411 722,clip]{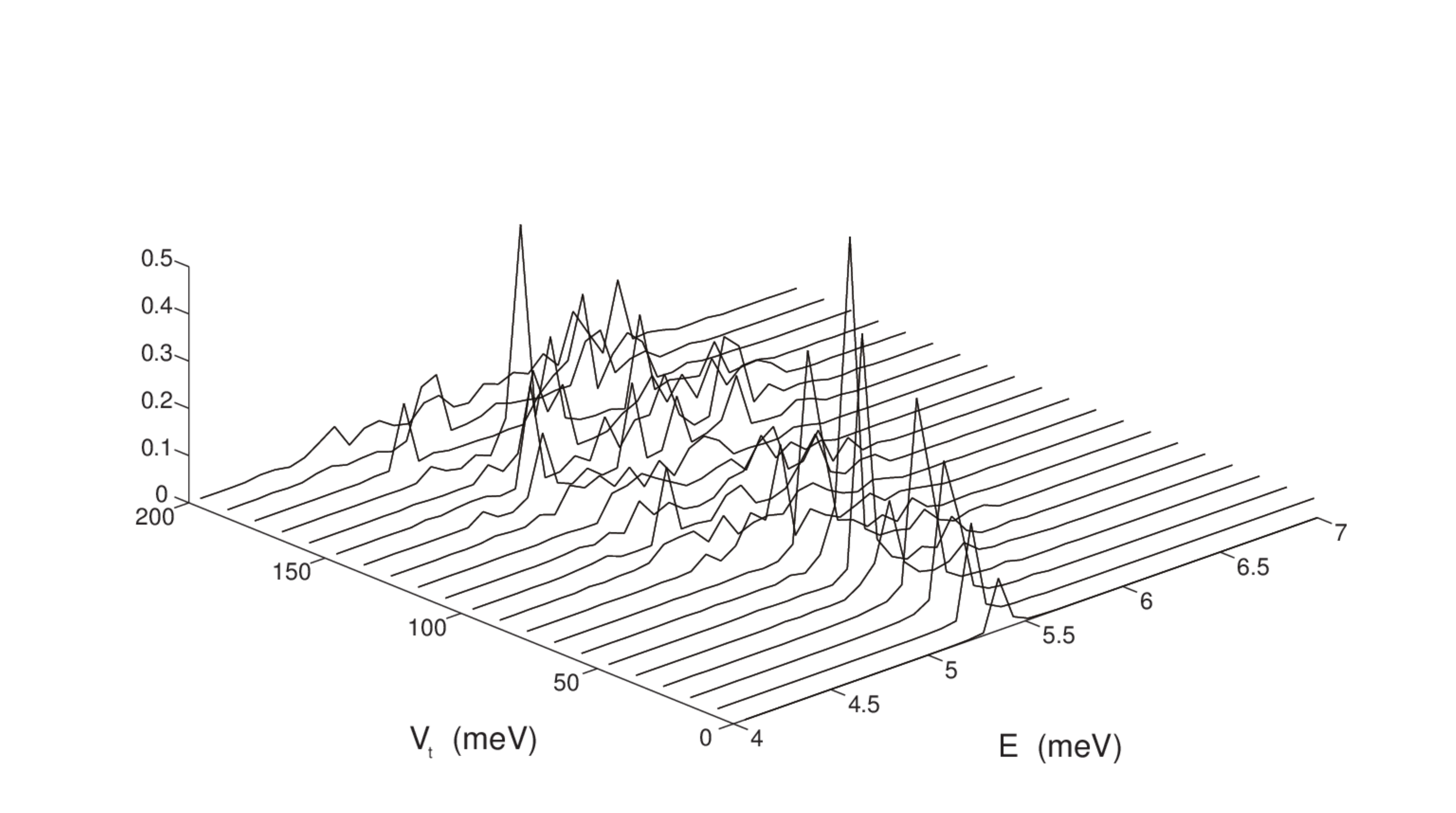}
      \caption{(Color online) The time-dependent expectation value of 
              $\langle r^2 \rangle$ and the corresponding Fourier power spectrum for the 
              Schr{\"o}dinger-Poisson model of the quantum dot with a central hill. $V_0=3.0$ meV, $T=0$ K.}
      \label{FFT-Schr-Poisson}
\end{figure}
In Fig.\ \ref{FFT-Schr-Poisson} we display the time-evolution of the expectation value 
$\langle\langle  \bar z \bar z^* \rangle\rangle$ and the corresponding Fourier
transform for the SP model with a central hill in the dot. 
Below, we will compare the location of the main peak
or peaks for low $V_t$ for the different models, but here we notice that the main
peak shows a local minimum around $V_t=40$ meV, a behavior not so different from the 
DFT model, but after $V_t=60$ meV the peak splits into a complex collection of smaller
peaks. For the system without a central hill (Fig.\ \ref{FFT-Schr-Poisson-nohill})
this disintegration of the main peaks happens earlier, and the resulting smaller
peaks are fewer than in the system with a central hill. 

The time-evolution in this essentially nonlinear model is very different from
what is known for linear models. In order to appreciate this fact better we look at a 
linear model before we comment futher on the time-evolution of the SP model.

\section{Exact time-evolution in a truncated Fock-space}
The CI-version of the model is capable to deliver the time-evolution of few
Coulomb interacting electrons in a quantum dot in an external magnetic field.
Here, we will use it for two electrons in the parabolic confinement introduced
earlier (\ref{Vpar}) with the option of the small central hill (\ref{Vc}).
The ground state for a vanishing external magnetic field is calculated in a
truncated two-particle Fock-space. The truncation limits the two-electron Fock-space
to the 16836 lowest states in energy. The Fock-space is constructed from the single-electron 
states of the parabolic confinement. The time evolution is again formally by the 
same Liouville-von Neuman equation (\ref{L-vN-dft}) as was used for the mean-field
version of the model, but now the density operator is a two-electron operator that
is expressed in the Fock-space for the interacting two electrons. 
The main difference here is that the Hamiltonian of the system is only time-dependent
as long as the initial perturbation (\ref{Wt}) is switched on. The Coulomb part of
the Hamiltonian is always time-independent and no iterations are necessary within
each time step in order to attain convergence for the interaction like in the case
of the DFT-model. 

The penalty of this approach is instead the size of the matrices
need for the calculation, but we have used two important technical items in order to
attain the time-evolution to 100 ps. First, we tested for the present parameters 
how much we could reduce the Fock-space for the time-integration of the time-evolution
operator (\ref{Teq}). The states which contribute for $V_t=200$ meV to the density matrix
with a contribution larger than $10^{-5}$ are less than 2415, so in the time-integration
we further truncate the Fock-space to that size. We remind that these 2415 interacting 
two-electron states were initially calculated using 16836 noninteracting two-electron
states. Still the matrices are considerably larger
than in the DFT-case, so we then rewrote the time-integration to run on powerful 
GPU's.\cite{Siro20121884} Furthermore, we tried two different 
methods for the time-integration, in one we refer the time-evolution operator
to the initial time $t=0$, and in the other one we only refer it to the one
earlier time step and accumulate the time-evolution in the density matrix.
We selected a time-step small enough for the methods to give the same
results.

\begin{figure}[htbq]
      \includegraphics[width=0.42\textwidth,angle=0]{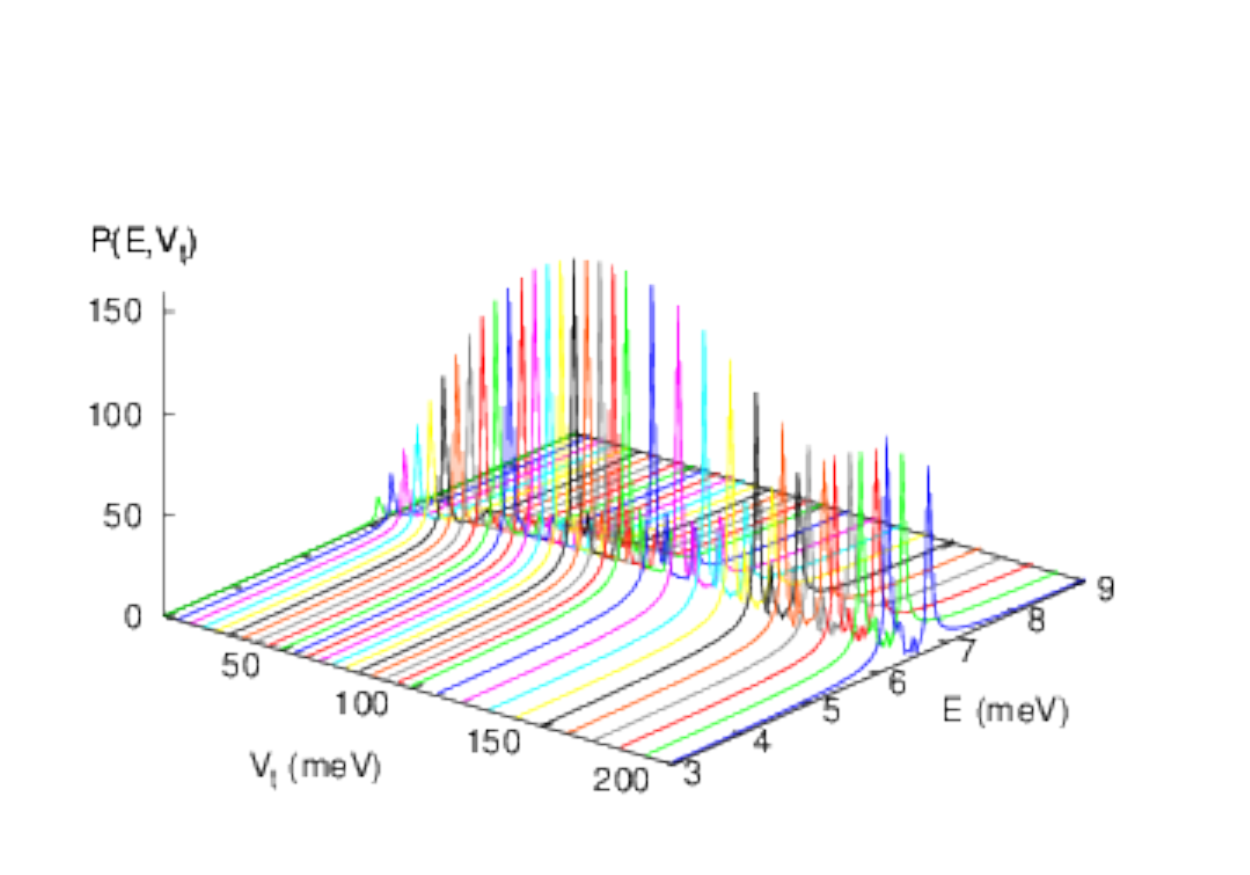}
      \includegraphics[width=0.42\textwidth,angle=0,bb=22 13 315 189,clip]{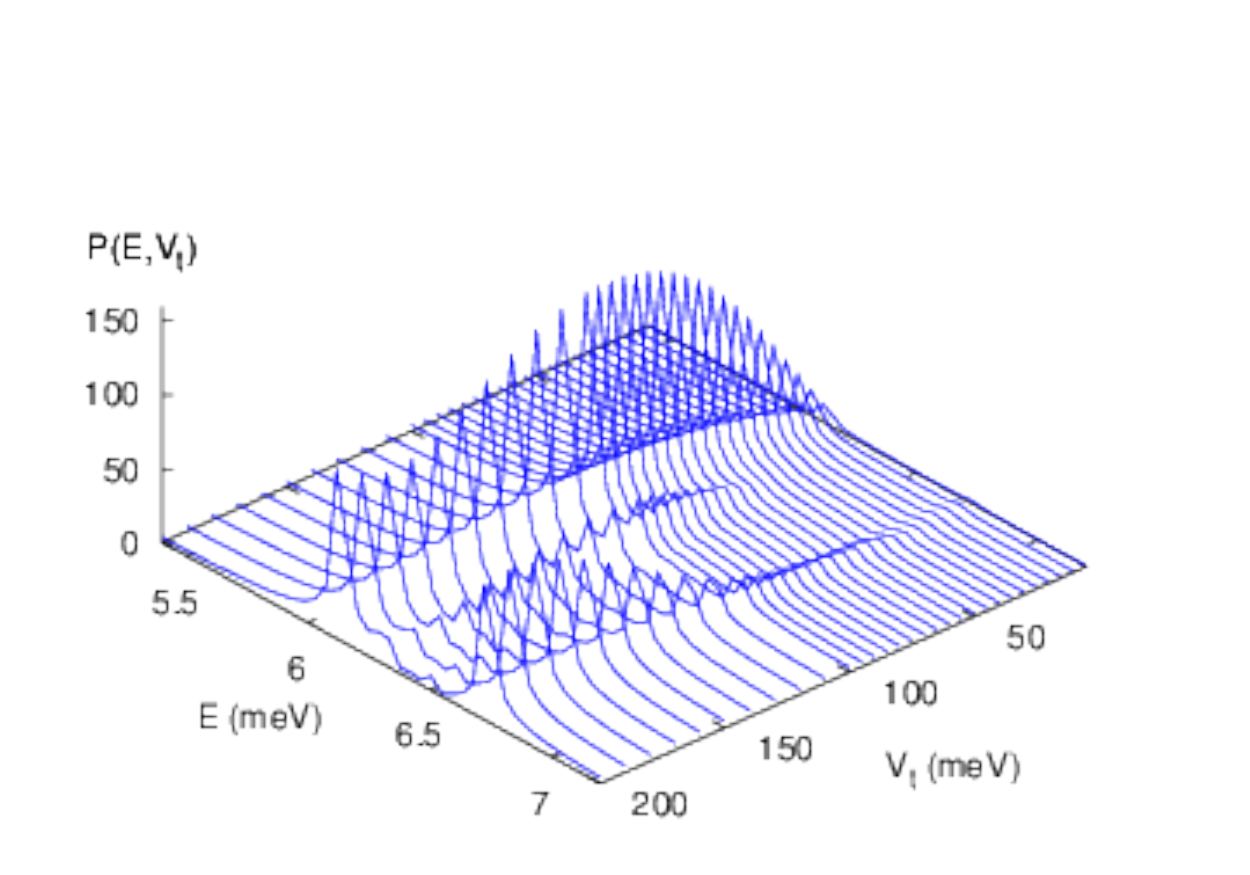}
      \caption{(Color online) The Fourier power spectrum for the time-dependent 
              expectation value of $\langle r^2\rangle$ for the CI model
              without a central hill. The lower panel focuses in on the
              energy axis close to resonances. $V_0=0$, $T=0$ K.}
      \label{FFT-exact}
\end{figure}

\begin{figure}[htbq]
      \includegraphics[width=0.42\textwidth,angle=0]{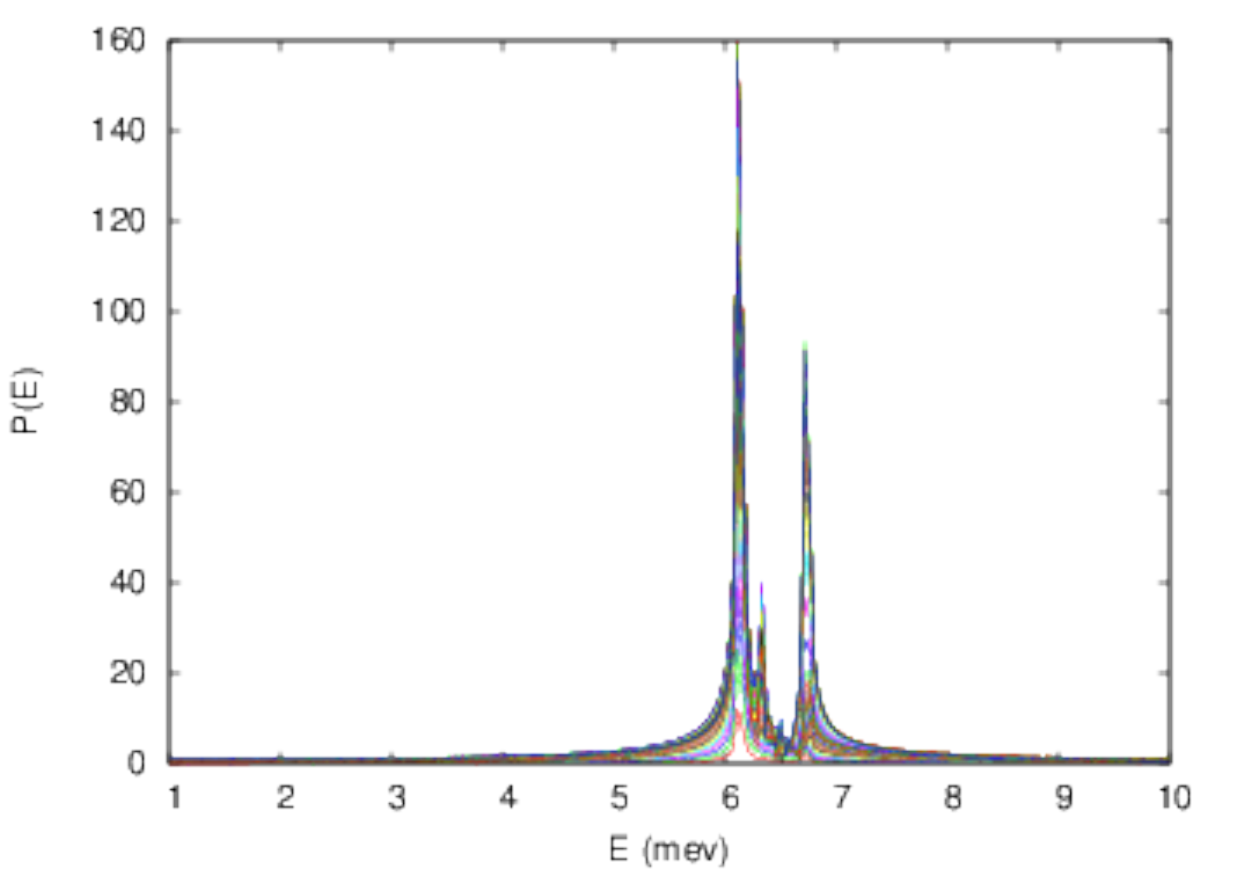}
      \includegraphics[width=0.42\textwidth,angle=0]{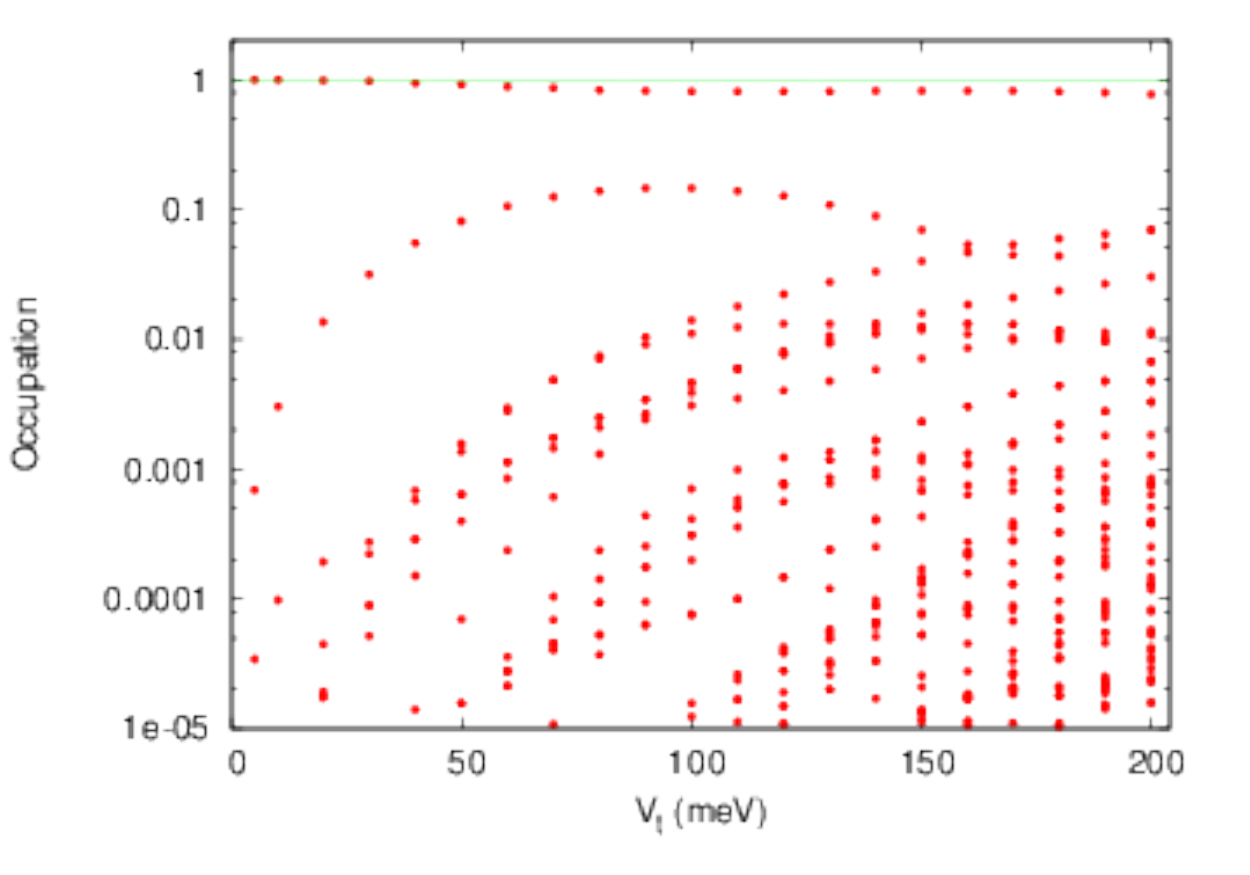}
      \caption{(Color online) The Fourier power spectrum for the time-dependent 
              expectation value of $\langle r^2\rangle$ for the CI model
              without a central hill (upper panel). The time-independent occupation of the 
              interacting two-electron states $|\alpha )$ after the perturbation pulse has 
              vanished (lower panel). $V_0=0$, $T=0$ K.}
      \label{FFT-exact-occ}
\end{figure}

After the initial perturbation pulse (\ref{Wt}) dies out nothing is explicitly
dependent on time in the Hamiltonian and therefore the diagonal elements of the
density matrix, the occupation of the interacting two-electron states stays constant.
In Figures \ref{FFT-exact} and \ref{FFT-exact-occ} we show the Fourier power spectrum for the collective
oscillations of the model expressed in terms of the expectation value $\langle r^2\rangle$, 
together with the time-independent occupation of each interacting two-electron state participating 
in the collective oscillations. Here, we have a pure
parabolic confinement without a central hill.

The logarithmic scale for the occupation in the lower panel of Fig.\ \ref{FFT-exact-occ}
hides the fact that for $V_t=200$ meV the occupation of the ground state has fallen to
77\%. This is another measure of the strength of the excitation.

\begin{figure}[htbq]
      \includegraphics[width=0.42\textwidth,angle=0]{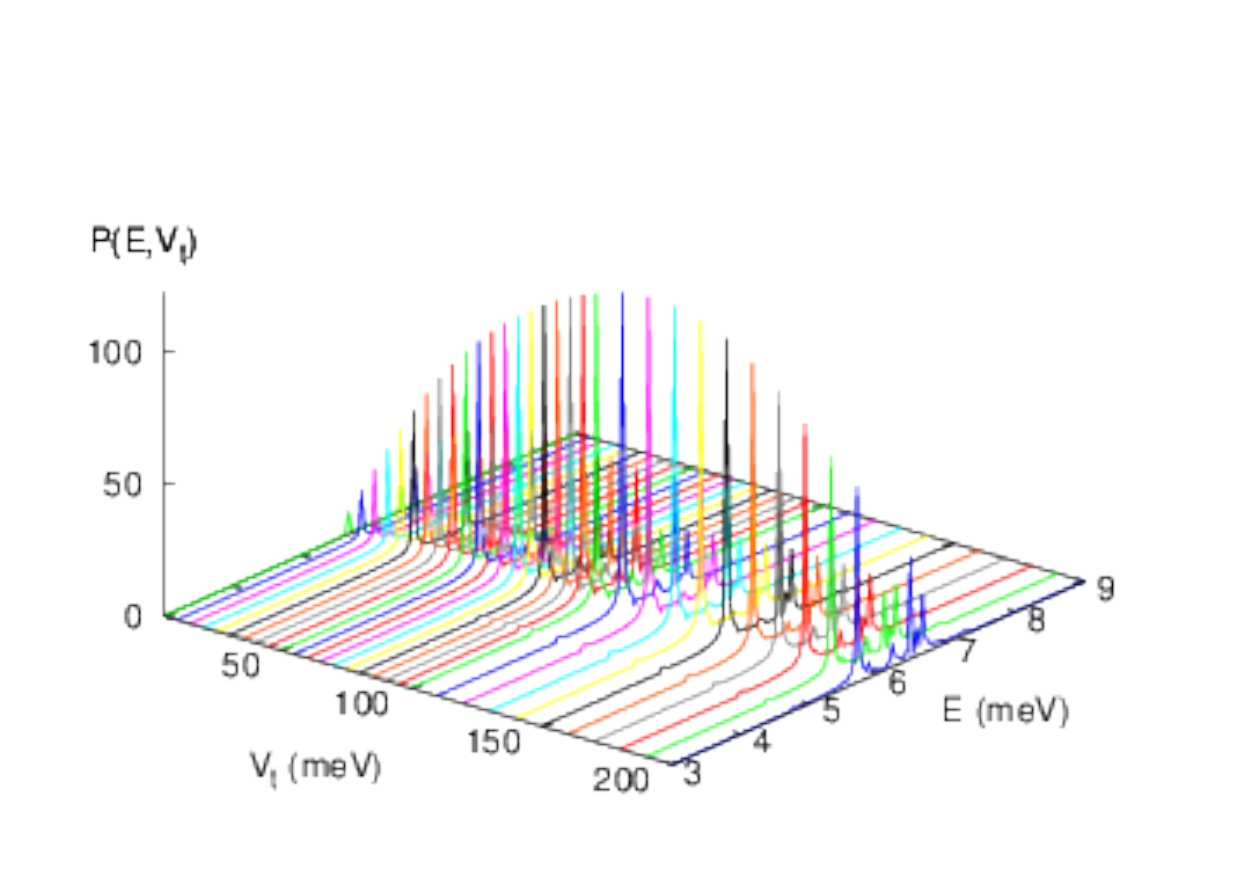}
      \includegraphics[width=0.42\textwidth,angle=0,bb=22 13 315 189,clip]{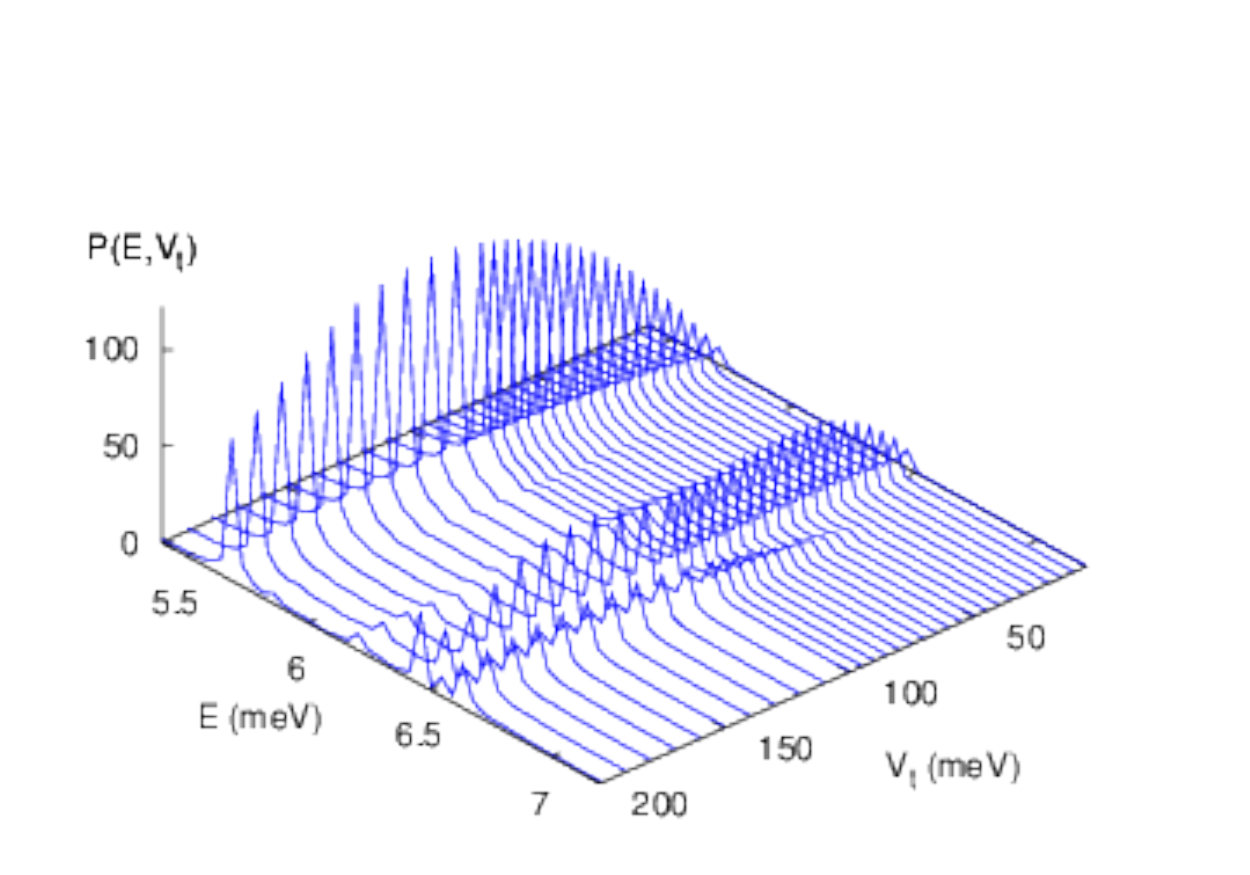}
      \caption{(Color online) The Fourier power spectrum for the time-dependent 
              expectation value of $\langle r^2\rangle$ for the CI model
              with a central hill. The lower panel focuses in on the
              energy axis close to resonances. $V_0=3.0$ meV, $T=0$ K.}
      \label{FFT-exact-hill}
\end{figure}

\begin{figure}[htbq]
      \includegraphics[width=0.42\textwidth,angle=0]{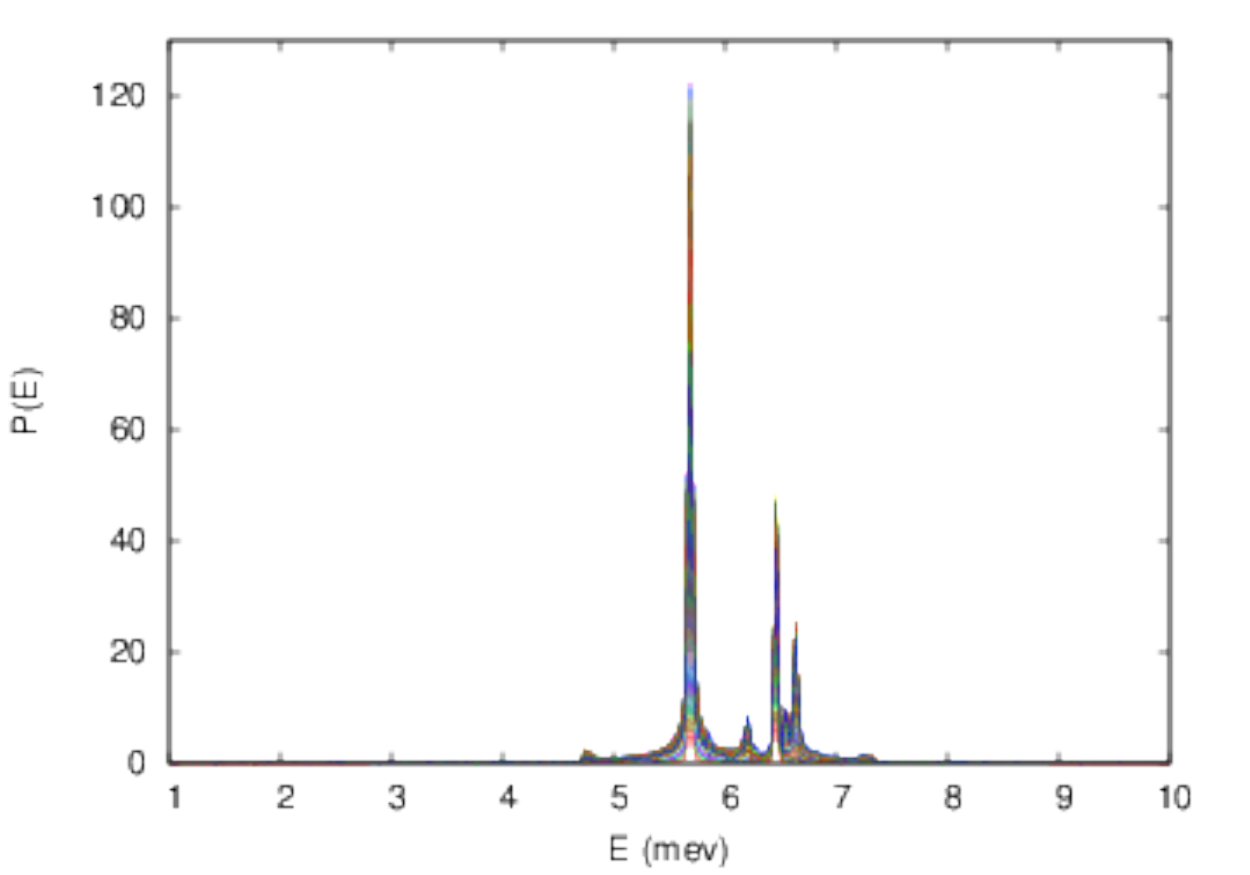}
      \includegraphics[width=0.42\textwidth,angle=0]{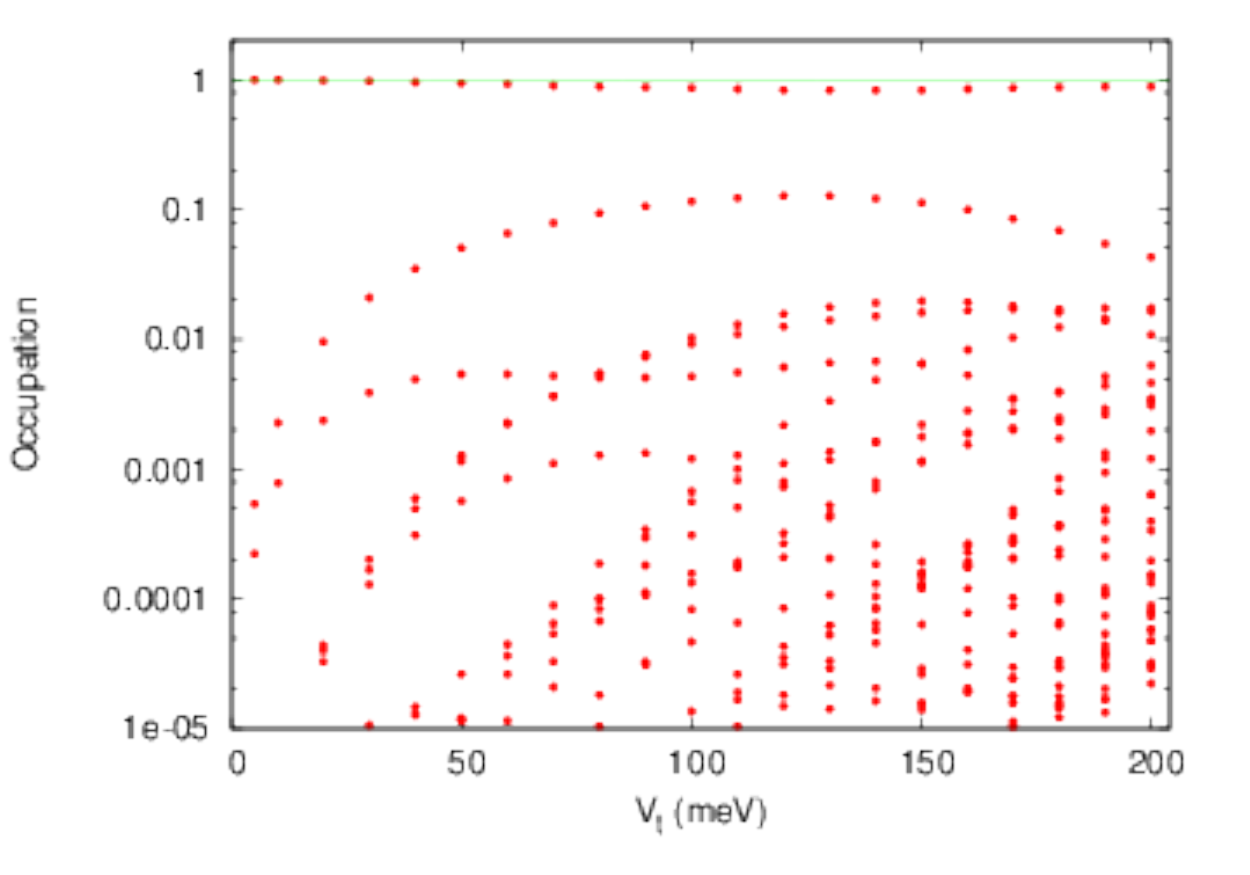}
      \caption{(Color online) The Fourier power spectrum for the time-dependent 
              expectation value of $\langle r^2\rangle$ for the CI model
              with a central hill (upper panel). The time-independent occupation of the 
              interacting two-electron states $|\alpha )$ after the perturbation pulse has 
              vanished (lower panel). $V_0=3.0$ meV, $T=0$ K.}
      \label{FFT-exact-hill-occ}
\end{figure}

The results for the quantum dot with a central hill (\ref{Vc}) added are shown in 
Figures \ref{FFT-exact-hill} and \ref{FFT-exact-hill-occ} 

The main surprise for the exact results is that we do not find any local minimum for
$V_t\approx 35-40$ meV. Indeed, the main peak found in the exact results shows behavior 
that is closer to the results of the HA if we consider only the height of the main peak
found. There are more peaks visible in the exact
results and that is reminiscent of the comparison in the linear response regime for the
exact and the Hartree-Fock approach.\cite{Pfannkuche94:1221} One might of course worry
about the possibility that the DFT-model could not predict the time-evolution properly
or could not describe the excited states correctly,
if it got stuck in some local minimum instead of a global minimum. We have tried to 
exclude this possibility by performing the DFT-calculation at higher temperatures,
$T=1.0$ and $4.0$ K. In both cases a minimum around $V_t\approx 35-40$ meV is found.
In addition, we have varied the minimum seeking, but in vain, the minimum always
reappears. 

The DFT-approach can be criticized by our use of a static functional instead of a
more appropriate frequency dependent one, especially since we are using it to  
describe a collective oscillation in the system. We have no good excuse for this, but
interestingly enough the DFT-model can reproduce the extended Kohn theorem valid
for parabolic confinement for $|N_p|=1$ with ease. The same test has of course been
used with success both for the exact CI-model and the Hartree-version of the DFT-model. 
Opposite to the CI-model the seeking of the ground state for the DFT model without
a central hill is a very time-consuming and difficult affair. This behavior has to
be related to the fact that the presence of the central hill (\ref{Vc}) reduces the
importance of the Coulomb interaction. In some sense this is also true for the 
nonphysical self-interaction in the Hartree-version of the DFT-model.

\begin{figure}[htbq]
      \includegraphics[width=0.42\textwidth,angle=0]{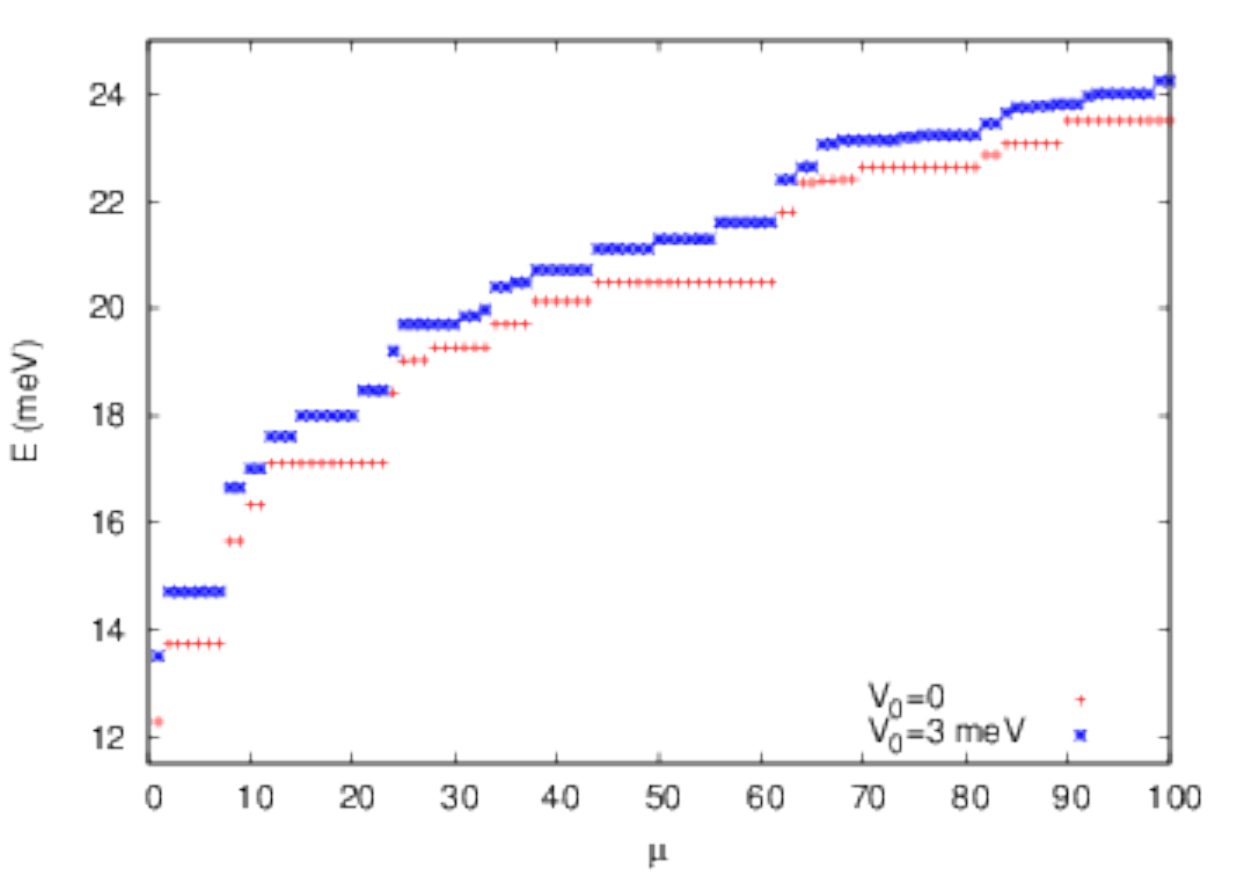}
      \includegraphics[width=0.42\textwidth,angle=0]{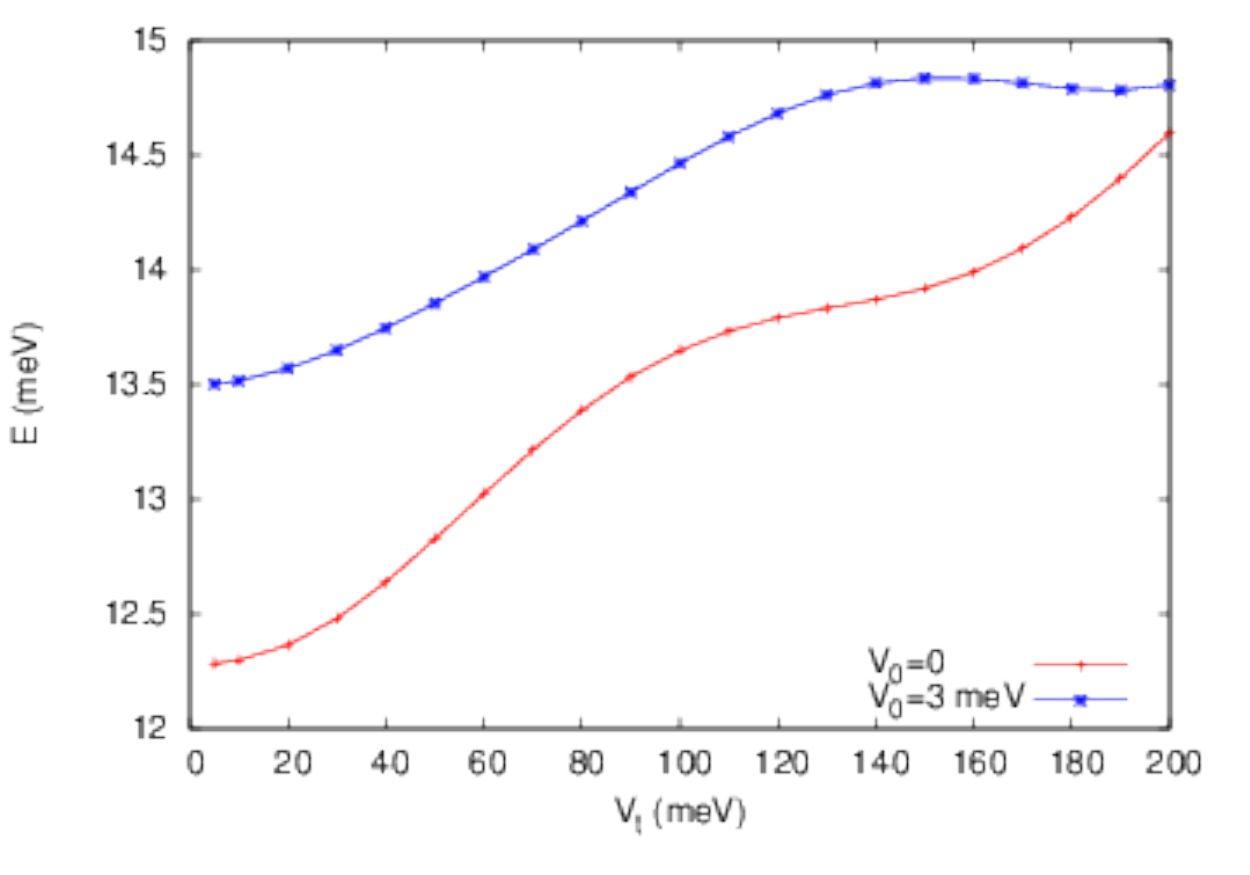}
      \caption{(Color online) The interacting two-electron spectra versus the 
               state number $\mu$ (upper panel), and total energy 
               versus the excitation strength $V_t$ (lower panel) compared for the exact model for the 
               system with ($V_0=3.0$ meV) and without ($V_0=0$) a central hill, $T=0$ K.}
      \label{Exact-E}
\end{figure}

Corresponding reduction of the importance of the Coulomb interaction in the case of the
CI-model can eventually be seen in the lower panel of Fig.\ \ref{FFT-exact-occ} and 
Fig.\ \ref{FFT-exact-hill-occ} for the occupation of the two-electron states caused
by the initial perturbation. The energy spectra for the 100 lowest interacting two-electron 
states are compared in the upper panel of Fig.\ \ref{Exact-E}. Besides the general behavior
of the central hill (\ref{Vc}) to increase the energy of each state we see a partial
lifting of degeneracy.  

We are here dealing with nonlinear response of a system as can be verified by looking at the 
expectation value for the total energy of the system described by the CI-model, after the 
excitation pulse has vanished, shown in Fig.\ \ref{Exact-E}. The excitation pulse pumps a
finite amount of energy into the system. This is important when interpreting
the occupation of the interacting two-electron states in the system displayed in the lower
panels of Fig.\ \ref{FFT-exact-occ} and \ref{FFT-exact-hill-occ}. If we look at the system without
a central hill, Fig.\ \ref{FFT-exact-occ}, we see that the ground state $|1)$ is occupied 
with probability close to 1, and for low excitation, $V_t$, the next state is $|24)$ and for 
higher $V_t$ state $|26)$ competes with $|24)$. If we check the energy differences we find
$E_{24}-E_1=6.139$ meV, and $E_{26}-E_1=6.746$ meV, which indeed fit with the main peak seen
and a side peak appearing for higher $V_t$ in Fig.\ \ref{FFT-exact}. 

For the case of a central hill in the system we find that again state $|24)$ has the next highest
occupation, but now for the whole $V_t$ range. Next comes state $|33)$ for low values of $V_t$.
Indeed, we get $E_{24}-E_1=5.698$ meV, and $E_{33}-E_1=6.472$ meV, which again fits very well with
the location of the peaks in Fig.\ \ref{FFT-exact-hill}. The graphs of the occupation of the 
interacting two-electron states $|\alpha )$ are thus indicating which states are being occupied 
as a result of the excitation of the system. We have verified that states $|24)$ and $|26)$ for the
system without a central hill and states $|24)$ and $|33)$ for the system with one, all have
a total angular momentum $\hbar{\cal M}=\hbar (M_1+M_2)=0$, where $M_i$ is the quantum number for
angular momentum of electron $i$. As was noted earlier\cite{Pfannkuche93:2244} the CI-model 
allows for contributions to an ${\cal M}=0$ state two single electron states with angular
momentum $\pm\hbar M$, a combination that is not possible in a HA with circular 
symmetry.\cite{Pfannkuche93:2244}  

\begin{figure}[htbq]
      \includegraphics[width=0.42\textwidth,angle=0]{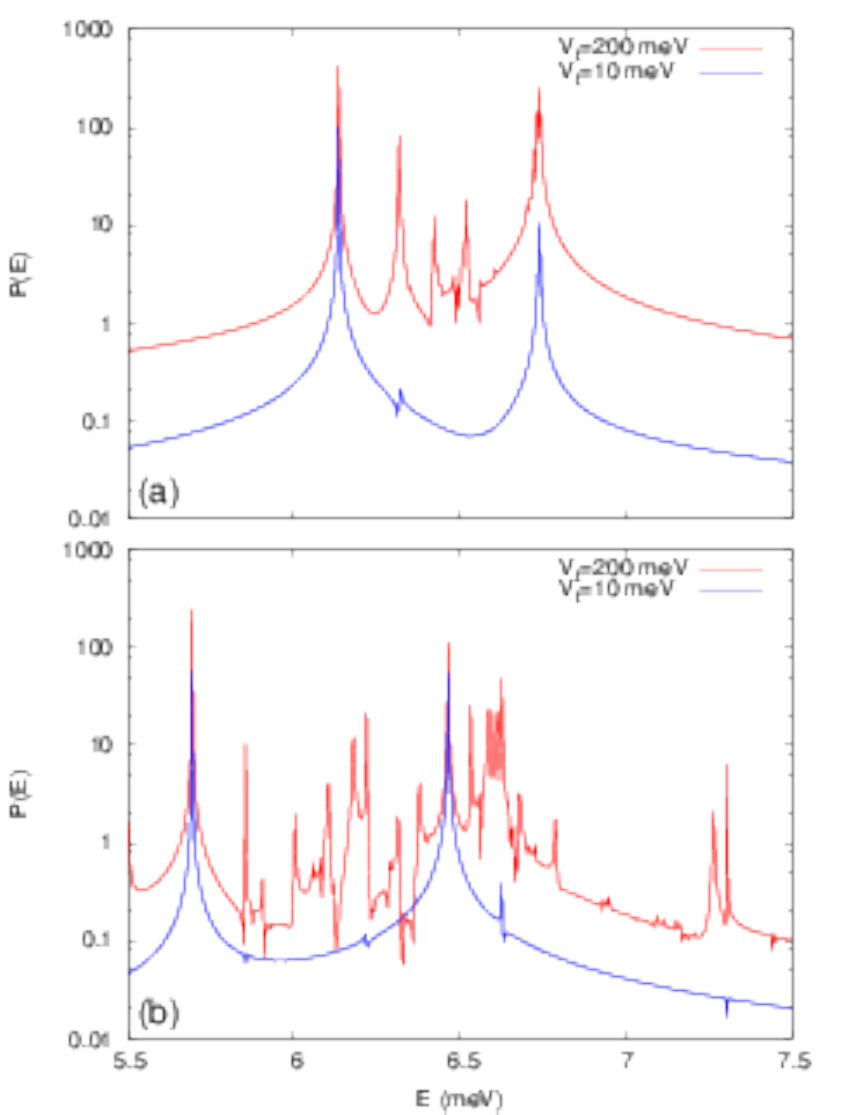}
      \caption{(Color online) The Fourier power spectra compared for $V_t=10$ meV and
              $V_t=200$ meV for the system without a central hill (a), and with a
              central hill (b). $T=0$ K.}
      \label{FFT-MaxMin}
\end{figure}

In Figure Fig.\ \ref{FFT-MaxMin} we compare the Fourier power spectra for 
$V_t=10$ meV and $V_t=200$ meV in the case of the system with a central hill 
and without one, but here we have taken an extra long time-series, integrating the 
equations of motion for 1000 ps instead of the 100 ps we have used for the CI-model
above.  
As could be expected for a linear model the peaks visible at low excitation are still
present with unchanged frequency for strong excitation, but the strong excitation activates 
several more peaks. It is also clear that the presence of the central hill (Fig.\ \ref{FFT-MaxMin}(b))
shifts the frequency of the main peaks and allows for the excitation of many more. 
The central hill does break some special symmetry imposed by the parabolic confinement, that
the Coulomb interaction alone does not break.

\section{Time-evolution of a Hubbard model}
Above, we have introduced mean-field theoretical models and a many-electron model
that is solved exactly in a truncated Fock-space for two electrons to describe the 
strong radial excitation
of electrons in a quantum dot. These models do all appear in different studies of linear
response of quantum dots. The mean-field models tend, due to their nature, though to be
used for dots with a higher number of electrons. The nonlinear SP-model can though be
considered as an attempt to create a version of a mean-field approach fit for two
electrons. For curiosity we like to add the last model, the Hubbard model, a many-electron
model that has not often been applied to describe the electrons in a single parabolically 
confined quantum dot.  

The Hamiltonian for the electrons in a quantum dot described by the Hubbard model is
\begin{equation}
    H = H_{\mathrm{int}} + H_{\mathrm{hop}} + H_{\mathrm{V}},
\label{Hub-H}
\end{equation}
where the Coulomb interaction between the electrons is described by a 
spin dependent contact interaction
\begin{equation}
    H_{\mathrm{int}} = U \sum_{i=1}^N  n_{i,\downarrow} n_{i,\uparrow},
\label{Hub-int}
\end{equation}
and the hopping part has the form
\begin{equation}
    H_{\mathrm{hop}} = -t \sum_{\sigma=\downarrow,\uparrow} \sum_{\langle i,j\rangle}
                       c_{i,\sigma}^\dagger c_{j,\sigma} +h.c.,
\label{Hub-hop}
\end{equation}
where $\langle i,j\rangle$ denotes a summation over the neighboring sites.
The model is written in terms of the creation $c_{i,\sigma}^\dagger$, the destruction $c_{i,\sigma}$, 
and the number operator $n_{i,\sigma}$ for electrons with spin $\sigma$ on site $i$. 
The potential part 
\begin{equation}
    H_\mathrm{V} = \sum_{\sigma=\downarrow,\uparrow} \sum_{i=1}^N V(r_i) n_{i,\sigma},
\label{Hub-Vpar}
\end{equation}
includes the parabolic potential (\ref{Vpar}) and possibly the central small hill (\ref{Vc}).

We set the Hubbard model on a small square lattice with totally $N$ sites.
We use the numbering of the states in the Fock-space suggested by Siro and Harju.\cite{Siro20121884}
The height and the width of the lattice is fixed in terms of the characteristic length scale for
the parabolic confinement to be $6a$. The lattice length is then $a_{\mathrm{latt}}=6a/(\sqrt{N}-1)$
and the hopping constant is $t=\hbar^2/(2m^*a_\mathrm{latt})$. The value for the strength of the 
Coulomb interaction is not so straightforward to find, but we fix the value of $U$ such that the 
energy of the ground state of the system is in accordance with the value found in the exact model.  
We keep in mind that there will always be a difference in the many-body energy spectrum of these 
two models, due to the different treatment of the Coulomb interaction and the finite square lattice
that is bound to break the angular symmetry of the original model, but we want to see if we can
identify some many-electron character in the excitation response. 

\begin{figure}[htbq]
      \includegraphics[width=0.42\textwidth,angle=0]{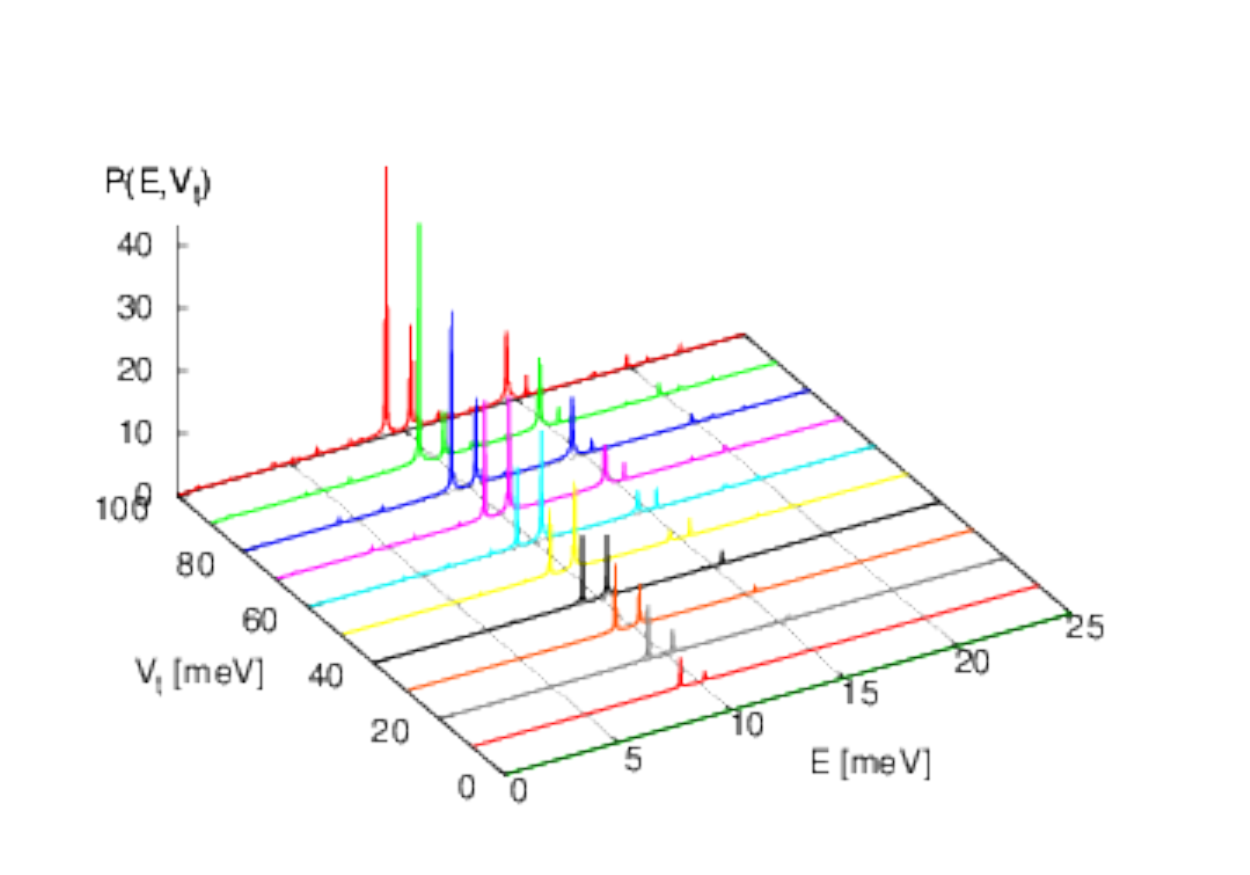}
      \includegraphics[width=0.42\textwidth,angle=0]{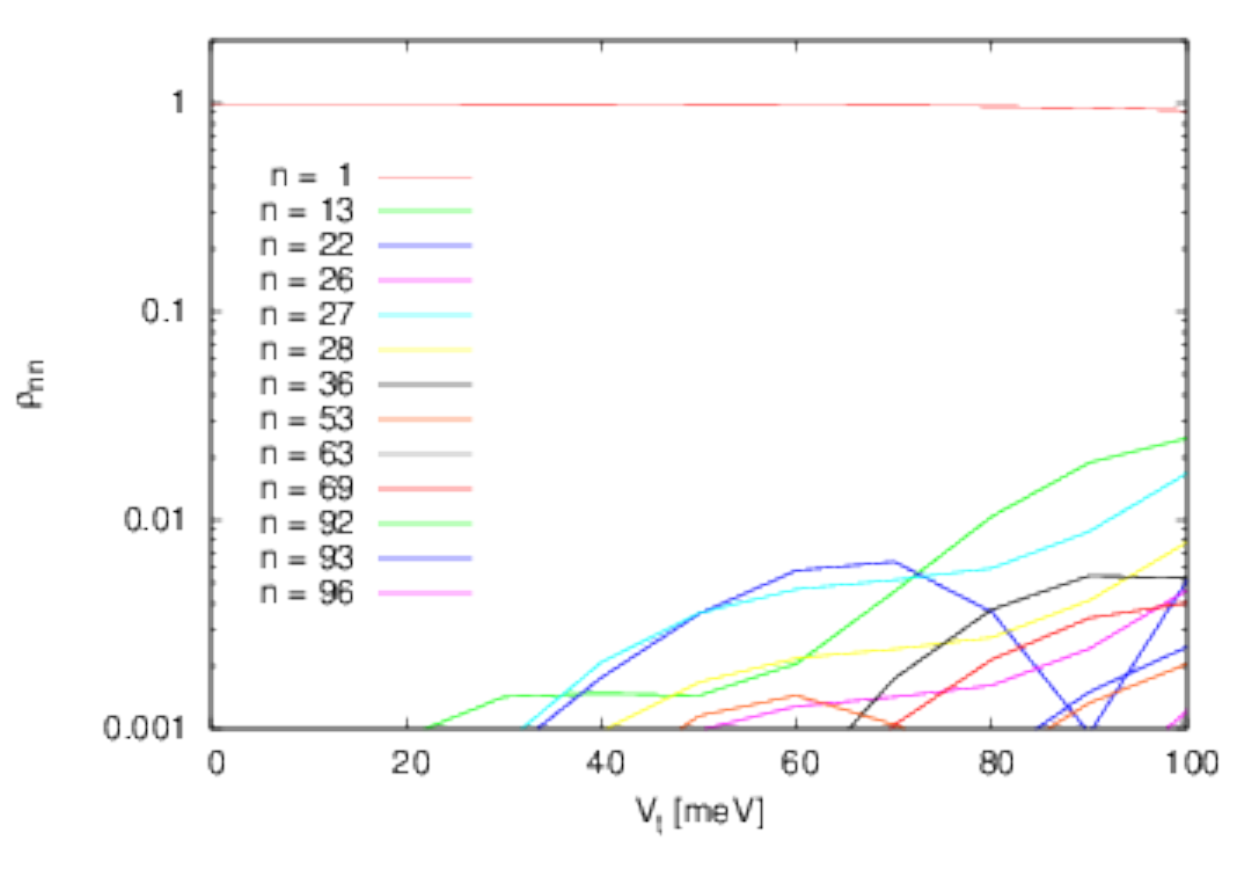}
      \caption{(Color online) The Fourier power spectrum of the expectation value of 
              $\langle r^2\rangle$ (upper panel), and the occupation of the interacting
              two-electron states (lower panel) for the Hubbard model without
              a central hill. $V_0=0$, $T=0$ K.}
      \label{FFT-hubbard}
\end{figure}

\begin{figure}[htbq]
      \includegraphics[width=0.42\textwidth,angle=0]{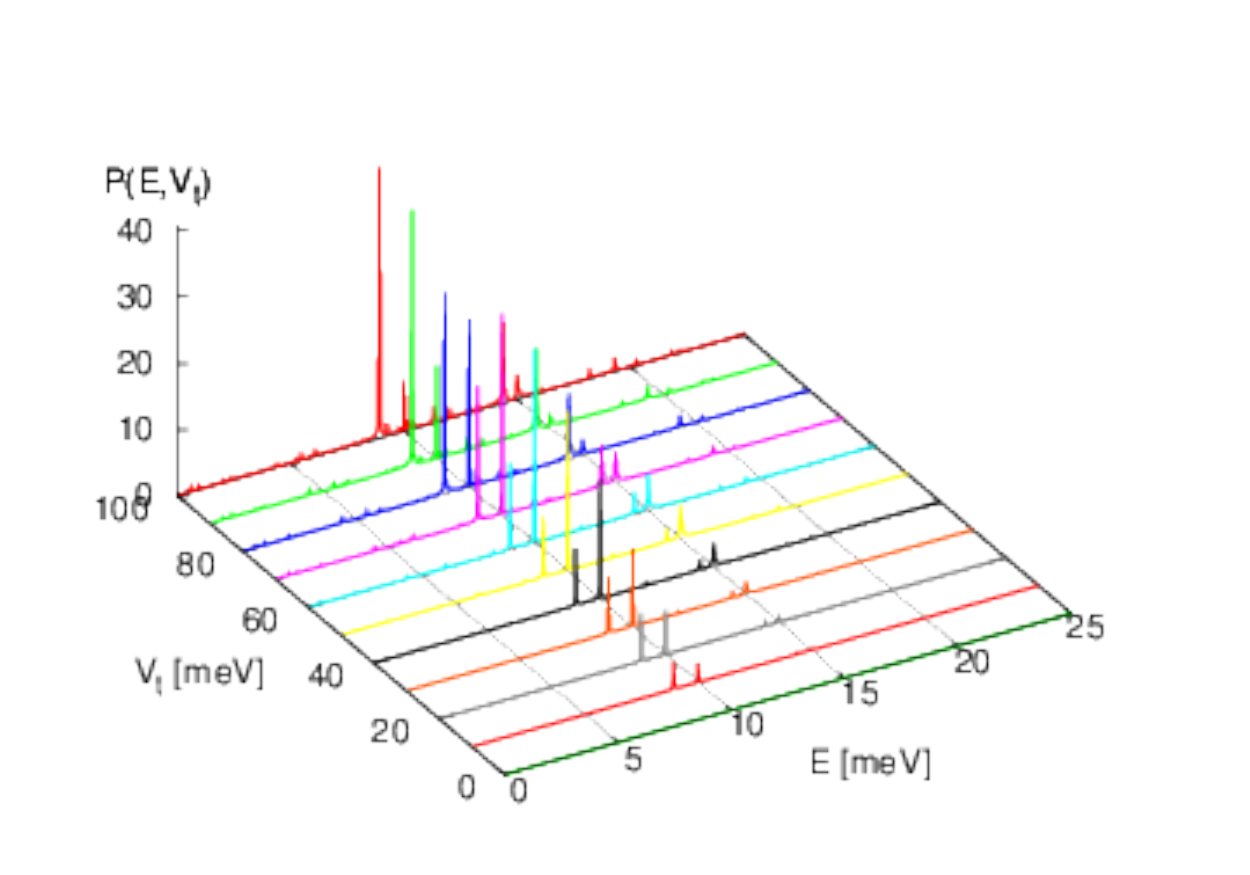}
      \includegraphics[width=0.42\textwidth,angle=0]{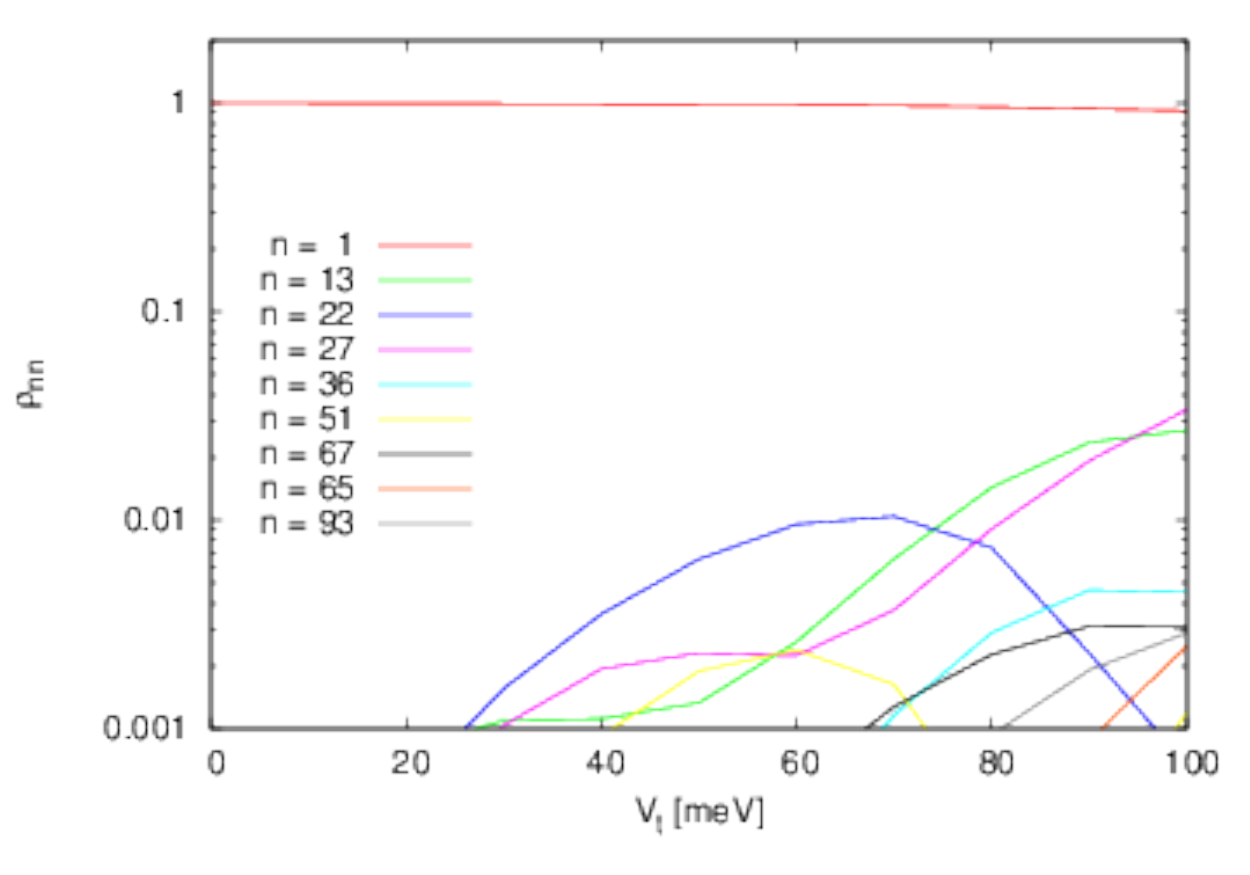}
      \caption{(Color online) The Fourier power spectrum of the expectation value of 
              $\langle r^2\rangle$ (upper panel), and the occupation of the interacting
              two-electron states (lower panel) for the Hubbard model with
              a central hill. $V_0=3.0$ meV, $T=0$ K.}
      \label{FFT-hubbard-hill}
\end{figure}

The parabolic confinement of the electrons spreads out the energy spectrum of the Hubbard model 
that otherwise is extremely dense, and thus we can use the same approach as for the exact many-body
model to solve it exactly within a truncated Fock-space. We performed this on GPU's for a 
$5\times 5$ lattice. The results for the Fourier transform of the 
time-dependent oscillations in $\langle r^2\rangle$
are shown in Fig.\ \ref{FFT-hubbard} for the pure parabolic confinement, and in Fig.\ \ref{FFT-hubbard-hill}
for the model with a small central hill (\ref{Vc}).

The time-evolution of the system is calculated in the same way as was used 
for the CI model using the time-evolution operators presented above (\ref{Teq})
in an interacting two-electron basis.

The square symmetry of the lattice can be expected to produce deviations 
that should already be present in the excitation spectrum for low  
excitation.\cite{PhysRevB.60.16591} We have tested the Hubbard model for dipole
active excitation modes, $N_p=\pm 1$, to verify this.  
The main peak (the lowest excitation) is indeed split for the Hubbard model 
in Figures \ref{FFT-hubbard} and \ref{FFT-hubbard-hill},
and the modes at higher energy, only appear for a stronger excitation,
i.e.\ a higher value of $V_t$. By looking at the lower panels in Figures
\ref{FFT-hubbard} and \ref{FFT-hubbard-hill} we see again that in the 
system without a central hill (Fig.\ \ref{FFT-hubbard}) more modes get active as
the excitation grows. This is in accordance with our observation for the CI model.
We have to admit that on this small lattice chosen the energy of the lowest mode
is a bit higher than all the other models predict, even though we have chosen the
interaction strength $U$ to give the similar energy for the ground state as the
CI model does. We do not use the Hubbard model for higher excitation than
$V_t=100$ meV to avoid artifacts created by the finite size of the model.

\section{Comparison of model results and discussion}
For the quantum dot with no central hill present ($V_0=0$) at weak excitation,
$V_t\sim 0$, all the models deliver one main peak that grows linearly with the 
excitation strength. As we have seen the Hubbard model due to the square symmetry
imposed by the underlying lattice has two peaks,\cite{PhysRevB.60.16591} and in the case of the CI model
we see a small side peak on the ``blue'' side, reminiscent of known results for the 
$N_p=\pm 1$ modes.\cite{Pfannkuche94:1221} 

The location of the main peak in the case of the SP model is redshifted by an amount
slightly surpassing 0.5 meV. This must be accredited to the slightly weaker repulsion 
of the electrons having a logarithmic singularity in the case of the SP model instead
of the 3D Coulomb repulsion in the other mean-field approximations, and in the CI model.
With the small central hill in the quantum dot the location of the main peak in the DFT 
model is blue shifted by 0.2 meV compared to the CI model, and the same analysis gives 
a blueshift of 0.3 meV for the HA. Calculations of the ground state for the CI, the SP,
and the DFT model all give similar energy, but the HA gives results far off.\cite{Pfannkuche93:2244} 

In the CI, the SP, and the Hubbard models the inclusion of a small central potential
hill in the confinement potential of the quantum dot causes more collective modes to
be activated with increasing excitation $V_t$. At the same time the lower panels of 
Figures \ref{FFT-exact-occ}, \ref{FFT-exact-hill-occ}, \ref{FFT-hubbard}, and 
\ref{FFT-hubbard-hill} displaying occupation of interacting two-electron states 
indicates a slight simplification effects caused by the central hill, at least
for some range of $V_t$. Amazingly, in the HA only one peak for the collective
oscillations is seen for the whole range of excitation strength we try.
This is probably caused by the artificial self-interaction that is specially
large for two electrons described with the HA. On the other hand the dependence 
of the height of the main peak on $V_t$ for the Fourier transform of the expectation value
$\langle r^2\rangle$ for the HA is very close to the results for the main peak for the 
CI model.   

The finite occupation of higher energy states together with the 
increase of the mean total energy seen in the lower panel of Fig.\ \ref{Exact-E}
shows that we have left the linear response regime with increasing $V_t$.          
This fact is further demonstrated by the nonlinear growth of the height of the Fourier peak for 
the expectation value $\langle r^2\rangle$ for all models with increasing $V_t$ beyond 
the linear regime for low $V_t$. In addition, we notice that for low $V_t$ the 
electrons in the quantum dot oscillate with $\langle r^2\rangle$ very close to the 
ground state value. As the excitation is increased energy is pumped into the quantum 
dot and it increases in size.

We have identified nonlinear behavior in all the models when observing how the 
amplitude of the oscillations of $\langle r^2\rangle$ behave as a function of
the excitation strength $V_t$ once we leave the linear response regime valid
for very low excitation. The CI model is a purely linear model. All the possible
excited states for the CI model are calculated before the time-integration of the 
system is started. This is clearly demonstrated in Fig.\ \ref{FFT-MaxMin} where the 
excitations are compared for weak and strong excitation. The main peak at low $V_t$ is 
still visible in the excitation spectrum for large $V_t$, at exactly the same energy.
Stronger excitation activates higher lying collective modes, and even in this simple
system there very are many of them available.  

The time-evolution for the mean-field models has to be viewed in different terms.
In case of the DFT or the Hartree model information about the two-electron excitation 
spectrum does not exist before the time-integration is started. The effective potential
changes in each time-step and the occupation of effective single-electron states becomes 
time-dependent, see Fig.\ \ref{Occupation-T}.   
The effective potential (or equivalently, the density, or the density operator here) has to
be found by iterations in each time-step in order to include the effects of the Coulomb
interaction. Within each iteration the problem is treated as a linear one. In case of the
SP model the nonlinear solution for the groundstate is sought directly without an iteration, 
and the same is true for the time-dependent solutions. The time-evolution of the SP model
is thus nontrivial and could in principle bring forward phenomena that could be blocked by
the linear solution requirement within each iteration step for the other mean-field models,
especially in a long time series where small effects from this methodology gathered in each
time-step might sum up.   

Within the range for $V_t$ considered here the HA brings results that look very stable, 
one peak with no frequency shift as $V_t$ increases, but with a slight nonlinear behavior for
the amplitude of the oscillations of $\langle r^2\rangle$. The SP and the DFT models bring
similar results for $V_t\leq 60$ meV with a local minimum for the amplitude of the oscillations
of $\langle r^2\rangle$. For larger values of $V_t$ the SP model brings a plethora of collective
oscillations, and for the DFT model it becomes too difficult to stabilize a solution for a longer
time interval. It should be kept in mind that also the CI model, especially for the case of no
central hill, shows an increased number of active modes, but only for much stronger excitation and
in a more ``controlled'' way. The different characteristics or the nuance of the nonlinear 
properties of the mean-field models may be influencing their response here to a strong excitation 
in a fundamentally different way than in the linear CI model.

It should be stated once more that extreme care has been taken in verifying and 
testing our numerical results by comparing different numerical methods, models, and
variation of sizes and types of functional spaces and grids.  

\section{Conclusion}
The modeling of nonlinear response of confined quantum systems on the nanoscale
is in its infancy, but may bring new insight into the systems as the measuring,
processing, and growth techniques evolve opening up the field. For systems with many 
particles we most likely will have to rely on mean-field and DFT models, and
only for systems of few particles can we expect to be able to rely on CI models.
In anticipation of this we have studied here how some of these models fare 
describing the nonlinear response of a two electron model.   

We have to expect the CI-model to deliver numerically exact results that we
can compare the results of the other models to. The results of our implementation 
of a DFT-model do not compare well when leaving the linear response regime.
This is not totally unexpected as we have not used any time-dependent functionals.
In addition, the numerical time-integration of the DFT model is difficult to guarantee
for strong excitation and long times. The Hartree model is easier to use and the overall
qualitative nonlinear response of it is in accordance with the CI model, except for fine
structure of side peaks visible in the CI model. Similar comparison has been seen in the linear
response earlier \cite{Pfannkuche94:1221}. The results of the coarse lattice Hubbard
model deviate quantitatively from the CI-results, but the qualitative behavior is similar,
side peaks and occupation of higher modes with increased excitation.

Regarding the emergence of nonlinear effects the comparison to the SP model is
valuable. In a mean field, or a local approximation to a DFT theory the results are
usually obtained by iterations, and most often there is a condition that the underlying
linear basis is orthogonal. In calculations of molecules this condition is sometimes relaxed, 
but most often it is used to guarantee a connection to higher order many-body methods.
This is not done in the SP model. There the nonlinear solution is found directly and
the resulting states are not orthogonal. Looking at our results we see that this 
essential nonlinearity does strongly affect the solution of the SP model beyond some
excitation strength. These effects, emergence of many new excitation modes, splitting of
modes, is not seen in any of the other models. So, even if the mean-field and the DFT
models are nonlinear, then the iteration procedure in a linear functional space does 
protect them from this mode splitting and multiplication. As stated in the previous 
section the nonlinear behavior seen from the CI results is much more modest and probably
only results from the ``shape'' of the many-body energy spectrum that can be reached with
increasing excitation. 

All these points in the end only stress how exciting and important experimental undertaking into 
this nonlinear regime will be.     

%
%
%
\begin{acknowledgments}
      This work was supported by the Research Fund of the University of Iceland, 
      the Icelandic Research and Instruments Funds, and a Special Initiative for Students of the 
      Directorate of Labour. Some of the calculations
      were performed on resources provided by the Nordic High Performance Computing (NHPC).
      C.\ Besse and G.\ Dujardin are partially supported by the French
      programme Labex CEMPI (ANR-11-LABX-0007-01).
\end{acknowledgments}
%
%
%
\bibliographystyle{apsrev4-1}
%
%
%
%
\end{document}